\documentclass[11pt,draft,onecolumn,final]{./IEEEtran}
\usepackage{amssymb,subfigure,amsthm,mathtools,enumerate}
\usepackage[dvips]{graphicx,color} %use for latex->dvips
\usepackage{fullpage}
\usepackage{cite}

\newcommand{\R}{\mathbb{R}}

\newcommand{\N}{\mathbb{N}}
\newcommand{\A}{\mathcal{A}}
\newcommand{\U}{\mathcal{U}}
\newcommand{\sC}{\mathcal{C}}
\newcommand{\uv}{\mathbf{u}}
\newcommand{\e}{\mathbf{e}}
\newcommand{\sd}{\Sigma\Delta}

\newtheorem{thm}{Theorem}%[section]
\newtheorem{prop}[thm]{Proposition}
\newtheorem{cor}[thm]{Corollary}  
\newtheorem{lm}[thm]{Lemma}

\theoremstyle{definition}
\newtheorem{defn}{Definition}%[section]
\theoremstyle{remark}

\title{The Golden Ratio Encoder}
\author{Ingrid~Daubechies,
  C.~Sinan~G{\"u}nt{\"u}rk, %Courant Institute of Mathematical Sciences \\
  Yang~Wang, % Michigan State University
  and~\"{O}zg\"{u}r~Y{\i}lmaz%
  \thanks{Ingrid Daubechies is with the Department of Mathematics and
    with the Program in Applied and Computational Mathematics,
    Princeton University, Princeton, NJ 08544 USA
    (email:ingrid@math.princeton.edu).}%
  \thanks{C.~Sinan~G{\"u}nt{\"u}rk is with the Courant Institute of
    Mathematical Sciences, New York, NY 10012 USA (email:gunturk@courant.nyu.edu).}%
\thanks{Yang Wang is with the Department of
    Mathematics, Michigan State University, East Lansing, MI 48824 USA
    (email:ywang@math.msu.edu).}%
  \thanks{\"{O}zg\"{u}r Y{\i}lmaz is with the Department of
    Mathematics, The University of British Columbia, Vancouver, BC V6T
    1Z2 Canada (email:oyilmaz@math.ubc.ca).}%
} %University of British Columbia}
\begin{document}
\maketitle

\begin{abstract} 
  This paper proposes a novel Nyquist-rate analog-to-digital (A/D)
  conversion algorithm which achieves exponential accuracy in the
  bit-rate despite using imperfect components.  The proposed algorithm
  is based on a robust implementation of a {\em beta-encoder} with
  $\beta=\phi=(1+\sqrt{5})/2$, the {\em golden ratio}.  It was
  previously shown that beta-encoders can be implemented in such a way
  that their exponential accuracy is robust against threshold offsets
  in the quantizer element.  This paper extends this result by
  allowing for imperfect analog multipliers with imprecise gain values
  as well. We also propose a formal computational model for
  algorithmic encoders and a general test bed for evaluating their
  robustness.

% In this paper, we propose a new Nyquist-rate analog-to-digital (A/D)
%   conversion algorithm which achieves exponential precision in the
%   bit-rate despite using imperfect components.  Our method is based on
%   a robust implementation of a {\em beta-encoder} with
%   $\beta=\phi=(1+\sqrt{5})/2$, the {\em golden ratio}.  It was shown
%   previously that beta-encoders can be implemented in such a way that
%   their exponential accuracy is robust against threshold offsets in
%   the quantizer element.  This paper extends this result by allowing
%   for imperfect analog multipliers with imprecise gain values as
%   well. We also propose a formal computational model for algorithmic
%   encoders and a general test bed for evaluating their robustness.

\end{abstract}
\IEEEpeerreviewmaketitle
\section{Introduction} 
% Digital computers and their efficiency in processing data is one of
% the main driving forces behind much of our modern technology.  On the
% other hand, in many occasions, the data sets we would like to store
% and process, e.g., images, audio, and video, are analog by their
% nature. This places a high demand on providing accurate conversion
% between the analog and digital worlds. The technology used in
% the analog-to-digital conversion, by necessity, involves analog
% devices which have physical limitations that, at first sight, conflict
% with accuracy demands. 
% In this paper, we shall introduce a new A/D
% conversion scheme, which we call the ``Golden Ratio Encoder'', that
% can be implemented robustly on analog hardware and enjoys an
% exponentially precise conversion performance.

In A/D conversion, the aim is to quantize analog signals, i.e., to
represent analog signals, which take their values in the continuum, by
finite bitstreams. Basic examples of analog signals include audio
signals and natural images. Frequently, the signal is first sampled on
a grid in its domain, which is sufficiently dense so that perfect (or
near-perfect) recovery from the acquired sample values is ensured by
an appropriate sampling theorem.  After this so-called sampling stage,
there are two common strategies to quantize the sequence of sample
values. {\em Oversampling} or $\sd$ Analog-to-Digital converters (or
ADCs) incorporate memory elements in their structure so that the
quantized value of a sample depends on other sample values and their
quantization. The accuracy of such converters can be assessed by
comparing the continuous input signal with the continuous output
signal obtained from the quantized sequence (after a digital-to-analog
(D/A) conversion stage).  In contrast, {\em Nyquist-rate} ADCs
quantize sample values separately, with the goal of approximating each
sample value as closely as possible using a given bit-budget, i.e.,
the number of bits one is allowed to use to quantize each sample. In
the case of Nyquist-rate ADCs, the analog objects of interest reduce
to real numbers in some interval, say $[0,1]$.

In this paper we focus on Nyquist-rate ADCs. Let $x \in [0,1]$. The
goal is to represent $x$ by a finite bitstream, say, of length $N$. A
straightforward approach is to consider the standard binary (base-2)
representation of $x$,
\begin{equation} \label{bin_exp}
x=\sum_{n=1}^\infty b_n 2^{-n}, \ \ b_n\in \{0,1\}.
\end{equation}
and to let $x_N$ be the $N$-bit truncation of the infinite series in
\eqref{bin_exp}, i.e.,
\begin{equation} \label{bin_exp_trun}
x_N=\sum_{n=1}^N b_n 2^{-n}.
\end{equation}
It is easy to see that $|x-x_N|\le 2^{-N}$ so that
$(b_1,b_2,\dots,b_N)$ provide an $N$-bit quantization of $x$ with
\emph{distortion} not more than $2^{-N}$. This method is known as
\emph{pulse code modulation (PCM)} and essentially provides the most
efficient encoding in a rate-distortion sense.

As our goal is analog-to-digital conversion, the next natural question
is how to compute the bits $b_n$ on an analog circuit. One popular
method to obtain $b_n$, called \emph{successive approximation},
extracts the bits using a recursive operation. Let $x_0=0$ and suppose
$x_n$ is as in \eqref{bin_exp_trun} for $n\ge 1$. Define $u_n :=
2^n(x-x_n)$ for $n \ge 0$. It is easy to see that the sequence
$(u_n)_0^\infty$ satisfies the recurrence relation
\begin{equation} \label{base-2-rec}
u_n = 2 u_{n-1} - b_n, ~~~ n=1,2,\dots,
\end{equation}
and the bits can simply be extracted via the formula 
\begin{equation} \label{b_n-formula}
b_n = \lfloor 2 u_{n-1} \rfloor = \left \{
\begin{array}{ll}
1, & u_{n-1} \geq 1/2, \\
0, & u_{n-1} < 1/2. \\
\end{array}
\right. 
\end{equation} 
Note that the relation (\ref{base-2-rec}) is
the \emph{doubling map}  in disguise; $u_n = T(u_{n-1})$, where
$T: u \mapsto 2u \;(\mbox{mod } 1)$.

The successive approximation algorithm as presented above provides an
algorithmic circuit implementation that computes the bits $b_n$ in the
binary expansion of $x$ while keeping all quantities ($u_n$ and $b_n$)
macroscopic and bounded, which means that these quantities can be held
as realistic and measurable electric charges. However, despite the
fact that base-2 representations, which essentially provide the
optimal encoding in the rate-distortion sense, can be computed via a
circuit, they are not the most popular choice of A/D conversion
method. This is mainly because of robustness concerns: In practice,
analog circuits are never precise, suffering from arithmetic errors
(e.g., through nonlinearity) as well as from quantizer errors (e.g.,
threshold offset), simultaneously being subject to thermal noise. All
relations hold only approximately, and therefore, all quantities are
approximately equal to their theoretical values. In the case of the
algorithm described in \eqref{base-2-rec} and \eqref{b_n-formula},
this means that the approximation will exceed acceptable error bounds
after only a finite (small) number of iterations because the dynamics
of an expanding map has ``sensitive dependence on initial
conditions''. This central problem in A/D conversion (as well as in
D/A conversion) has led to the development of many alternative bit
representations of numbers, as well as of signals, that have been
adopted and/or implemented in circuit engineering, such as
beta-representations and $\Sigma\Delta$ modulation.

From a theoretical point of view, the imprecision problem associated
to the successive approximation algorithm is not a sufficient reason
to discard the base-$2$ representation out of hand. After all, we do
not have to use the specific algorithm in \eqref{base-2-rec} and
\eqref{b_n-formula} to extract the bits, and conceivably there could
be better, i.e., more resilient, algorithms to evaluate $b_n(x)$ for
each $x$. However, the root of the real problem lies deeper: the bits
in the base-$2$ representation are essentially uniquely determined,
and are ultimately computed by a greedy method. Since $2^{-n} =
2^{-n-1}+2^{-n-2}+\dots$, there exists essentially no choice other
than to set $b_n$ according to \eqref{b_n-formula}. (A possible choice
exists for $x$ of the form $x=k 2^{-m}$, with $k$ an odd integer, and
even then, only the bit $b_m$ can be chosen freely: for the choice
$b_m=0$, one has $b_n=1$ for all $n>m$; for the choice $b_m=1$, one
has $b_n=0$ for all $n>m$.) It is clear that there is no way to
recover from an erroneous bit computation: if the value $1$ is
assigned to $b_n$ even though $x < x_{n-1} + 2^{-n}$, then this causes
an ``overshoot'' from which there is no way to ``back up''
later. Similarly assigning the value $0$ to $b_n$ when $x > x_{n-1} +
2^{-n}$ implies a ``fall-behind'' from which there is no way to
``catch up'' later.

Due to this lack of robustness, the base-$2$ representation is not the
preferred quantization method for A/D conversion.  For similar
reasons, it is also generally not the preferred method for D/A
conversion.  In practical settings, oversampled coarse quantization
($\Sigma\Delta$ modulation) is more popular \cite{IY,CT,ST},
mostly due to its robustness achieved with the help of the redundant
set of output codes that can represent each source value \cite{GLV}.
However, standard $\Sigma\Delta$ modulation is suboptimal as a
quantization method (even though exponential accuracy in the bit rate
can be achieved \cite{exp_decay}). 

A partial remedy comes with fractional base expansions, called
$\beta$-representations
\cite{DK,DDGV,DY,KLB,P}.  Fix $1 < \beta <
2$. It is well known that every $x$ in $[0,1]$ (in fact, in
$[0,(\beta-1)^{-1}]$) can be represented by an infinite series
\begin{equation} \label{base-beta-rep}
x = \sum_{n=1}^{\infty} b_n \beta^{-n}, 
\end{equation}
with an appropriate choice of the bit sequence $(b_n)$. Such a
sequence can be obtained via the following modified successive
approximation algorithm: Define $u_n = \beta^n(x - x_n)$. Then the
bits $b_n$ obtained via the recursion
\begin{eqnarray} \label{b_n-formula-beta}
u_n &=& \beta u_{n-1} - b_n, \nonumber \\
b_n  &=& \left \{
\begin{array}{ll}
0, & u_{n-1} \leq a, \\
0 \mbox{ or } 1, & u_{n-1} \in (a,b), \\
1, & u_{n-1} \geq b. \\
\end{array}
\right. 
\end{eqnarray}
satisfy \eqref{base-beta-rep} whenever $1/\beta \le a \le b \le
1/\beta(\beta{-}1)$. If $a=b=1/\beta$, the above recursion is called
the \emph{greedy selection algorithm}; if $a=b= 1/\beta(\beta{-}1)$ it
is the \emph{lazy selection algorithm}. The intermediate cases
correspond to what we call \emph{cautious selection}. An immediate
observation is that many distinct $\beta$-representations in the form
(\ref{base-beta-rep}) are now available. In fact, it is known that for
any $1<\beta < 2$, almost all numbers (in the Lebesgue measure sense)
have uncountably many distinct $\beta$-representations \cite{S1}.
Although $N$-bit truncated $\beta$-representations are only accurate
to within $O(\beta^{-N})$, which is inferior to the accuracy of a
base-$2$ representation, the redundancy of $\beta$-representation
makes it an appealing alternative since it is possible to recover from
(certain) incorrect bit computations. In particular, if these mistakes
result from an unknown threshold offset in the quantizer, then it
turns out that a \emph{cautious} selection algorithm (rather than the
greedy or the lazy selection algorithms) is robust provided a bound
for the offset is known \cite{DDGV}. In other words, perfect encoding
is possible with an imperfect (flaky) quantizer whose threshold value
fluctuates in the interval $(1/\beta, 1/\beta(\beta-1))$.

It is important to note that a circuit that computes truncated
$\beta$-representations by implementing the recursion in
\eqref{b_n-formula-beta} has two critical parameters: the quantizer
threshold (which, in the case of the ``flaky quantizer'', is the pair
$(a,b)$) and the multiplier $\beta$. As discussed above,
$\beta$-encoders are robust with respect to the changes in the
quantizer threshold. On the other hand, they are \emph{not robust}
with respect to the value of $\beta$ (see Section \ref{dag} for a
discussion). A partial remedy for this has been proposed in \cite{DY}
which enables one to recover the value of $\beta$ with the required
accuracy, provided its value varies smoothly and slowly from one clock
cycle to the next.

In this paper, we introduce a novel ADC which we call \emph{the golden
  ratio encoder (GRE)}. GRE computes $\beta$-representations with
respect to base $\beta=\phi=(1+\sqrt{5})/2$ via an implementation that
is \emph{not} a successive approximation algorithm. We show that GRE
is robust with respect to its full parameter set while enjoying
exponential accuracy in the bit rate. To our knowledge, GRE is the
first example of such a scheme.

The outline of the paper is as follows: In Section \ref{enc_alg_rob1},
we introduce notation and basic terminology. In Section \ref{AE} we
review and formalize fundamental properties of \emph{algorithmic
  converters}. Section \ref{dag} introduces a computational model for
algorithmic converters, formally defines robustness for such
converters, and reviews, within the established framework, the
robustness properties of several algorithmic ADCs in the
literature. Section \ref{GRE} is devoted to GRE and its detailed
study. In particular, Sections \ref{GRE-1} and \ref{GRE-2} introduce
the algorithm underlying GRE and establish basic approximation error
estimates. In Section \ref{GRE-3}, we give our main result, i.e., we
prove that GRE is robust in its full parameter set. Sections
\ref{GRE-4}, \ref{GRE-5}, and \ref{GRE-6} discuss several additional
properties of GRE. Finally, in Section \ref{poly}, we comment on how
one can construct ``higher-order'' versions of GRE.

\section{Encoding, Algorithms and Robustness} \label{enc_alg_rob}
\subsection{Basic Notions for Encoding} \label{enc_alg_rob1}
We denote the space of analog objects to be quantized by $X$.  More
precisely, let $X$ be a compact metric space with metric
$d_X$. Typically $d_X$ will be derived from a norm $\| \cdot \|$
defined in an ambient vector space, via $d_X(x,y) = \|x-y\|$.  We say
that $E_N$ is an {\em $N$-bit encoder} for $X$ if $E_N$ maps $X$ to
$\{0,1\}^N$.  An infinite family of encoders $\{E_N\}_1^\infty$ is
said to be {\em progressive} if it is generated by a single map $E:X
\mapsto \{0,1\}^{\N}$ such that for $x\in X$,
\begin{equation} \label{encoderN}
E_N(x)=(E(x)_0, E(x)_1,\dots, E(x)_{N-1}).
\end{equation}
In this case, we will refer to $E$ as the generator, or sometimes
simply as the encoder, a term which we will also use to refer to the family
$\{E_N\}_1^\infty$.

We say that a map $D_N$ is a {\em decoder} for $E_N$ if $D_N$ maps the
range of $E_N$ to some subset of $X$. Once $X$ is an infinite set,
$E_N$ can never be one-to-one, hence analog-to-digital conversion is
inherently lossy. We define the {\em distortion} of a given
encoder-decoder pair $(E_N,D_N)$ by
\begin{equation} \label{distortion}
\delta_X(E_N,D_N) := \sup_{x\in X} ~ d_X(x,D_N(E_N(x))),
\end{equation}
and the {\em accuracy} of $E_N$ by
\begin{equation}\label{accuracy}
\A(E_N):=\inf_{D_N:\{0,1\}^N \to X} ~ \delta_X(E_N,D_N).
\end{equation}
The compactness of $X$ ensures that there exists a family of encoders
$\{E_N\}_1^\infty$ and a corresponding family of decoders
$\{D_N\}_1^\infty$ such that $ \delta_X(E_N,D_N) \to 0$ as $N \to
\infty$; i.e., all $x \in X$ can be recovered via the limit of
$D_N(E_N(x))$. In this case, we say that the family of encoders
$\{E_N\}_1^\infty$ is {\em invertible}. For a progressive family
generated by $E:X\to\{0,1\}^\N$, this actually implies that $E$ is
one-to-one.  Note, however, that the supremum over $x$ in
(\ref{distortion}) imposes uniformity of approximation, which is
slightly stronger than mere invertibility of $E$.

An important quality measure of an encoder is the rate at which
$\A(E_N) \to 0$ as $N\to \infty$. There is a limit to this rate which
is determined by the space $X$. (This rate is connected to the
Kolmogorov $\epsilon$-entropy of $X$, $\mathcal{H}_\epsilon(X)$,
defined to be the base-2 logarithm of the smallest number $k$ such
that there exists an $\epsilon$-net for $X$ of cardinality $k$
\cite{KT}. If we denote the map $\epsilon \mapsto
\mathcal{H}_\epsilon(X)$ by $\varphi$, i.e.,
$\varphi(\epsilon)=\mathcal{H}_\epsilon(X)$, then the infimum of
$\A(E_N)$ over all possible encoders $E_N$ is roughly equal to
$\varphi^{-1}(N)$.) Let us denote the number $\inf_{E_N:X\mapsto
  \{0,1\}^N} \A(E_N)$ by $\A_N(X)$. In general an optimal encoder may
be impractical, and a compromise is sought between optimality and
practicality. It is, however, desirable when designing an encoder that
its performance is close to optimal. We say that a given family of
encoders $\{E_N\}_1^\infty$ is {\em near-optimal for $X$}, if
\begin{equation} \label{near-optimal-X}
\A(E_N) \leq C \A_N(X),
\end{equation}
where $C \ge 1$ is a constant independent of $N$.  We will also say that
a given family of decoders $\{D_N\}_1^\infty$ is near-optimal for
$\{E_N\}_1^\infty$, if
\begin{equation} \label{near-optimal}
\delta_X (E_N,D_N) \leq C \A(E_N).
\end{equation}

An additional important performance criterion for an ADC is whether
the encoder is robust against perturbations. Roughly speaking, this
robustness corresponds to the requirement that for all encoders
$\{\widetilde E_N\}$ that are small (and mostly unknown) perturbations
of the original invertible family of encoders $\{E_N\}$, it is still
true that $\mathcal{A}(\widetilde E_N) \to 0$, possibly at the same
rate as $E_N$, using the same decoders. The magnitude of the
perturbations, however, need not be measured using the Hamming metric
on $\{0,1\}^X$ (e.g., in the form $\sup_{x \in X}
d_H(E_N(x),\widetilde E_N(x))$). It is more realistic to consider
perturbations that directly have to do with how these functions are
computed in an actual circuit, i.e., using small building blocks
(comparators, adders, etc.). It is often possible to associate a set
of internal parameters with such building blocks, which could be used
to define appropriate metrics for the perturbations affecting the
encoder.  From a mathematical point of view, all of these notions need
to be defined carefully and precisely.  For this purpose, we will
focus on a special class of encoders, so-called \emph{algorithmic
  converters}. We will further consider a computational model for
algorithmic converters and formally define the notion of robustness
for such converters.

\subsection{Algorithmic Converters} \label{AE}

\begin{figure}[t]
\begin{center}
\includegraphics[width=3in]{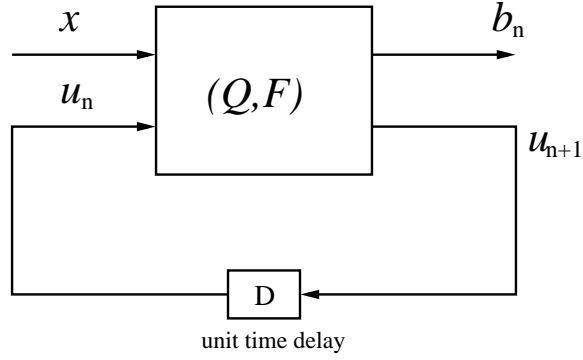} 
\end{center}
\caption{
\label{alg_enc}
The block diagram describing an algorithmic encoder.}
\end{figure}

By an {\em algorithmic converter}, we mean an encoder that can be
implemented by carrying out an autonomous operation (the algorithm)
iteratively to compute the bit representation of any input $x$.  Many
ADCs of practical interest, e.g., $\sd$ modulators, PCM, and
beta-encoders, are algorithmic encoders. Figure \ref{alg_enc} shows
the block diagram of a generic algorithmic encoder.

Let $\U$ denote the set of possible ``states'' of the converter
circuit that get updated after each iteration (clock cycle). More
precisely, let
$$Q: X \times \U \mapsto \{0,1\}$$ 
be a ``quantizer'' and let
$$F: X \times \U \mapsto \U$$ 
be the map that determines the state of the circuit in the next clock
cycle given its present state. After fixing the initial state of the
circuit $u_0$, the circuit employs the pair of functions $(Q,F)$ to
carry out the following iteration:
\begin{equation} \label{alg_eqn}
\left\{
\begin{array}{rcl}
b_n & = & Q(x,u_n) \\ 
u_{n+1} & = & F(x,u_n)
\end{array}
\right\},
 \ \ \ n=0,1,\dots
\end{equation}
This procedure naturally defines a progressive family of encoders
$\{E_N\}_1^\infty$ with the generator map $E$ given by
$$E(x)_n:=b_n,\ \ \ n=0,1,\dots$$
We will write $\text{AE}(Q,F)$ to refer to the algorithmic converter
defined by the pair $(Q,F)$ and the implicit initial condition $u_0$.
If the generator $E$ is invertible (on $E(X)$), then we say that the
converter is invertible as well.

\begin{defn}[1-bit quantizer] We define the 1-bit quantizer with
  threshold value $\tau$ to be the function
\begin{equation} \label{quant}
q_\tau(u) :=
\begin{cases}
0, & u<\tau, \\
1, & u \ge \tau.
\end{cases}
\end{equation}
\end{defn}

\subsubsection*{Examples of algorithmic encoders}
\begin{enumerate}
\item[\bf 1.]  {\em Successive approximation algorithm for PCM.}  The
  successive approximation algorithm sets $u_0 = x \in [0,2]$, and
  computes the bits $b_n$, $n=0,1,\dots$, in the binary expansion $x =
  \sum_0^\infty b_n 2^{-n} $ via the iteration
\begin{equation} \label{sa_pcm}
\left\{
\begin{array}{rcl}
b_n & = & q_1(u_n) \\ 
u_{n+1} & = & 2(u_n-b_n)
\end{array}
\right\},
 \ \ \ n=0,1,\dots
\end{equation}
Defining $Q(x,u)=q_1(u)$ and $F(x,u)=2(u-q_1(u))$, we obtain an
invertible algorithmic converter for $X=[0,2]$.  A priori we can set
$\U = \R$, though it is easily seen that all the $u_n$ remain in
$[0,2]$.

\item[\bf 2.]  {\em Beta-encoders with successive approximation
    implementation} \cite{DDGV}.  Let $\beta \in (1,2)$. A
  $\beta$-representation of $x$ is an expansion of the form $x =
  \sum_0^\infty b_n \beta^{-n}$, where $b_n \in \{0,1\}$.  Unlike a
  base-2 representation, almost every $x$ has infinitely many such
  representations. One class of expansions is found via the iteration
\begin{equation} \label{sa_beta}
\left\{
\begin{array}{rcl}
b_n & = & q_\tau(u_n) \\ 
u_{n+1} & = & \beta(u_n-b_n)
\end{array}
\right\},
 \ \ \ n=0,1,\dots,
\end{equation}
where $u_0 = x \in [0,\beta/(\beta-1)]$. The case $\tau = 1$ corresponds to the
``greedy'' expansion, and the case $\tau=1/(\beta-1)$ to the ``lazy''
expansion. All values of $\tau \in [1,1/(\beta-1)]$ are admissible in
the sense that the $u_n$ remain bounded, which guarantees the validity
of the inversion formula. Therefore the maps $Q(x,u)=q_\tau(u)$, and
$F(x,u)=\beta(u-q_\tau(u))$ define an invertible algorithmic encoder
for $X=[0,\beta]$.  It can be checked that all $u_n$ remain in a
bounded interval $\U$ independently of $\tau$.

\item[\bf 3.]  {\em First-order $\sd$ with constant
    input.} \label{ex_sd_1o} Let $x \in [0,1]$.  The first-order $\sd$
  (Sigma-Delta) ADC sets the value of $u_0 \in [0,1]$ arbitrarily and
  runs the following iteration:
\begin{equation} \label{sd_1o}
\left\{
\begin{array}{rcl}
b_n & = & q_1(u_n+x) \\ 
u_{n+1} & = & u_n + x - b_n
\end{array}
\right\},
 \ \ \ n=0,1,\dots
\end{equation}
It is easy to show that $u_n \in [0,1]$ for all $n$, and the
boundedness of $u_n$ implies
\begin{equation} \label{invert_sd}
x = \lim_{N \to \infty} \frac{1}{N}\sum_{n=1}^N b_n,
\end{equation}
that is, the corresponding generator map $E$ is invertible. For this
invertible algorithmic encoder, we have $X=[0,1]$, $\U = [0,1]$,
$Q(x,u)=q_1(u+x)$, and $F(x,u)=u+x-q_1(u+x)$.

\item[\bf 4.]  {\em $k$th-order $\sd$ with constant input.}  Let
  $\Delta$ be the forward difference operator defined by $(\Delta u)_n
  = u_{n+1}-u_n$. A $k$th-order $\sd$ ADC generalizes the scheme in
  Example 3 by replacing the first-order difference equation in
  \eqref{sd_1o} with
\begin{equation} \label{sd_k_diff1}
(\Delta^k u)_n = x - b_n, \ \ \ n=0,1,\dots,
\end{equation}
which can be rewritten as 
\begin{equation} \label{sd_k_diff2}
u_{n+k} = \sum_{j=0}^{k-1} a^{(k)}_j u_{n+j} + x - b_n, \ \ \ n=0,1,\dots,
\end{equation}
where $a^{(k)}_j = (-1)^{k-1-j} \binom{k}{j}$.  Here $b_n \in \{0,1\}$
is computed by a function $Q$ of $x$ and the previous $k$ state
variables $u_n,\dots,u_{n+k-1}$, which must guarantee that the $u_n$
remain in some bounded interval for all $n$, provided that the initial
conditions $u_0,\dots,u_{k-1}$ are picked appropriately. If we define
the vector state variable $\uv_{n} =\begin{bmatrix} u_n, \cdots,
  u_{n+k-1}\end{bmatrix}^\top$, then it is apparent that we can
rewrite the above equations in the form
\begin{equation} \label{sd_ko} 
\left\{
\begin{array}{rcl}
b_n & = & Q(x,\uv_n) \\ 
\uv_{n+1} & = & \mathbf{L}_k \uv_n + x\e - b_n\e
\end{array}
\right\},
 \ \ \ n=0,1,\dots,
\end{equation}
where $\mathbf{L}_k$ is the $k \times k$ companion matrix defined by
\begin{equation}
\mathbf{L}_k := 
\begin{bmatrix}
0 & 1 & 0 & \cdots & 0 \\
0 & 0 & 1 & \cdots & 0 \\
\vdots & \vdots & & \ddots & \vdots \\
0 & 0 & & \cdots & 1 \\
a^{(k)}_0 & a^{(k)}_1 & & \cdots & a^{(k)}_{k-1}
\end{bmatrix},
\end{equation}
and $\e = \begin{bmatrix} 0, \cdots, 0, 1 \end{bmatrix}^\top$.  If
$Q$ is such that there exists a set $\U \subset \R^{k}$ with the
property $\uv \in \U$ implies $F(x,\uv) := \mathbf{L}_k \uv+
(x-Q(x,\uv))\e \in \U$ for each $x \in X$, then it is guaranteed that
the $u_n$ are bounded, i.e., the scheme is {\em stable}. Note that any
stable $k$th order scheme is also a first order scheme with respect to
the state variable $(\Delta^{k-1}u)_n$. This implies that the
inversion formula (\ref{invert_sd}) holds, and therefore (\ref{sd_ko})
defines an invertible algorithmic encoder. Stable $\sd$ schemes of
arbitrary order $k$ have been devised in \cite{DD} and in
\cite{exp_decay}. For these schemes $X$ is a proper subinterval of
$[0,1]$.
\end{enumerate}

\subsection{A Computational Model for Algorithmic Encoders and 
Formal Robustness} \label{dag}

Next, we focus on a crucial property that is required for any ADC to
be implementable in practice. As mentioned before, any ADC must
perform certain arithmetic (computational) operations (e.g., addition,
multiplication), and Boolean operations (e.g., comparison of some
analog quantities with predetermined reference values). In the analog
world, these operations cannot be done with infinite precision due to
physical limitations. Therefore, the algorithm underlying a practical
ADC needs to be robust with respect to implementation imperfections.

In this section, we shall describe a computational model for
algorithmic encoders that includes all the examples discussed above
and provides us with a formal framework in which to investigate
others. This model will also allow us to formally define robustness
for this class of encoders, and make comparisons with the
state-of-the-art converters.

\subsubsection*{Directed Acyclic Graph Model}

Recall (\ref{alg_eqn}) which (along with Figure \ref{alg_enc})
describes one cycle of an algorithmic encoder.  So far, the pair
$(Q,F)$ of maps has been defined in a very general context and could
have arbitrary complexity.  In this section, we would like to propose
a more realistic computational model for these maps.  Our first
assumption will be that $X \subset \R$ and $\U \subset \R^d$.

\begin{figure}[t]
\begin{center}
\includegraphics[width=4in]{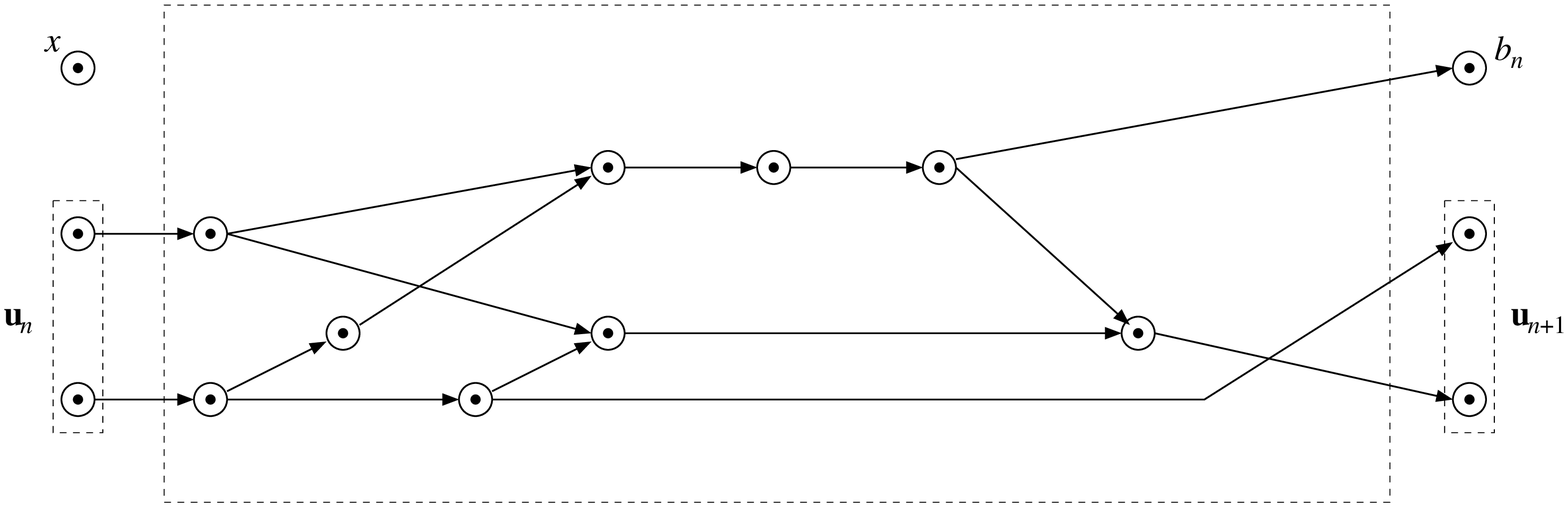} 
\end{center}
\caption{
\label{DAG}
A sample DAG model
for the function pair $(Q,F)$.}
\end{figure}

\begin{figure}[t]
\begin{center}
\includegraphics[width=6in]{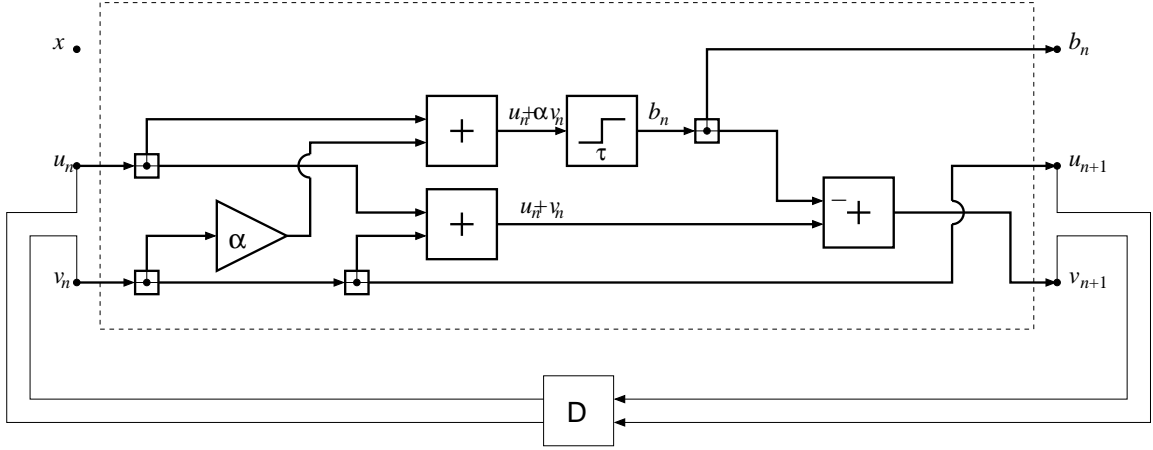} 
\end{center}
\caption{
\label{GRE_block}
Block diagram of the Golden Ratio Encoder (Section \ref{GRE})
corresponding to the DAG model in Figure \ref{DAG}.
}
\end{figure}

A {\em directed acyclic graph} (DAG) is a directed graph with no
directed cycles.  In a DAG, a {\em source} is a node (vertex) that has
no incoming edges. Similarly, a {\em sink} is a node with no outgoing
edges.  Every DAG has a set of sources and a set of sinks, and every
directed path starts from a source and ends at a sink.  Our DAG model
will always have $d+1$ source nodes that correspond to $x$ and the
$d$-dimensional vector $\uv_n$, and $d+1$ sink nodes that correspond
to $b_n$ and the $d$-dimensional vector $\uv_{n+1}$.  This is
illustrated in Figure \ref{DAG} which depicts the DAG model of an
implementation of the Golden Ratio Encoder (see Section \ref{GRE} and
Figure \ref{GRE_block}). Note that the feedback loop from Figure
\ref{alg_enc} is not a part of the DAG model, therefore the nodes
associated to $x$ and $\uv_n$ have no incoming edges in Figure
\ref{DAG}. Similarly the nodes associated to $b_n$ and $\uv_{n+1}$
have no outgoing edges. In addition, the node for $x$, even when $x$
is not actually used as an input to $(Q,F)$, will be considered only
as a source (and not as a sink).

\begin{figure}[t]
\begin{center}
\includegraphics[width=3.5in]{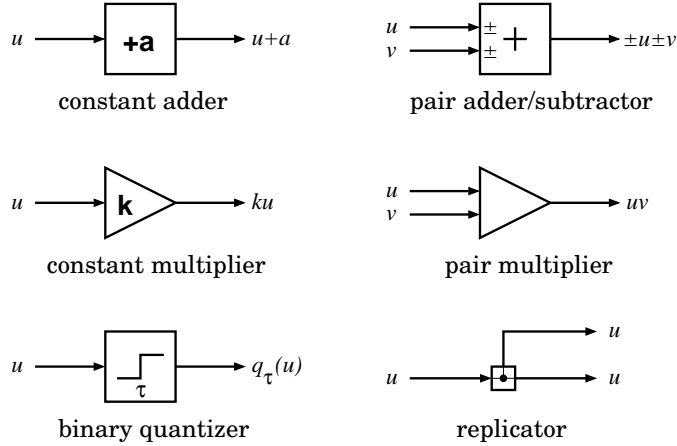} 
\end{center}
\caption{
\label{components}
A representative class of components in an algorithmic encoder.}
\end{figure}

In our DAG model, a node which is not a source or a sink will be
associated with a {\em component}, i.e., a computational device with
inputs and outputs.  Mathematically, a component is simply a function
(of smaller ``complexity'') selected from a given fixed class.  In the
setting of this paper, we shall be concerned with only a restricted
class of basic components: constant adder, pair adder/subtractor,
constant multiplier, pair multiplier, binary quantizer, and
replicator. These are depicted in Figure \ref{components} along with
their defining relations, except for the binary quantizer, which was
defined in (\ref{quant}). Note that a component and the node at which
it is placed should be consistent, i.e., the number of incoming edges
must match the number of inputs of the component, and the number of
outgoing edges must match the number of outputs of the component.

When the DAG model for an algorithmic encoder defined by the pair
$(Q,F)$ employs the ordered $k$-tuple of components $\sC =
(C_1,\dots,C_k)$ (including repetitions), we will denote this encoder
by $\text{AE}(Q,F,\sC)$.

\subsubsection*{Robustness}

We would like to say that an invertible algorithmic encoder
$\text{AE}(Q,F)$ is {\em robust} if $\text{AE}(\widetilde Q,
\widetilde F)$ is also invertible for any
$(\widetilde{Q},\widetilde{F})$ in a given neighborhood of $(Q,F)$ and
if the generator $E$ of $\text{AE}(Q,F)$ has an inverse $D$ defined on
$\{0,1\}^\N$ that is also an inverse for the generator $\widetilde E$
of $\text{AE}(\widetilde Q,\widetilde F)$.  To make sense of this
definition, we need to define the meaning of ``neighborhood'' of the
pair $(Q,F)$.

Let us consider an algorithmic encoder $\text{AE}(Q,F,\sC)$ as
described above. Each component $C_j$ in $\sC$ may incorporate a
vector $\lambda_j$ of parameters (allowing for the possibility of the
null vector). Let $\lambda = (\lambda_1,\dots,\lambda_k)$ be the
aggregate parameter vector of such an encoder. We will then denote
$\sC$ by $\sC(\lambda)$.

Let $\Lambda$ be a space of parameters for a given algorithmic encoder
and consider a metric $\rho_\Lambda$ on $\Lambda$.  We now say that
$\text{AE}(Q,F,\sC(\lambda_0))$ is {\em robust} if there exists a
$\delta > 0$ such that $\text{AE}(Q,F,\sC(\lambda))$ is invertible for
all $\lambda$ with $\rho_\Lambda(\lambda,\lambda_0) \leq \delta$ with
a common inverse. Similarly, we say that
$\text{AE}(Q,F,\sC(\lambda_0))$ is {\em robust in approximation} if
there exists a $\delta > 0$ and a family $\{D_N\}_1^\infty$ such that
$$ \delta_X(\widetilde E_N,D_N) \leq C \A(E_N)$$
for all $\widetilde E_N = \widetilde E_N(Q,F,\lambda)$ with $\rho_\Lambda(
\lambda,\lambda_0) \leq \delta$,  
where $E_N =E_N(Q,F,\lambda_0)$.

From a practical point of view, if an algorithmic encoder
$\text{AE}(Q,F,\sC(\lambda_0))$ is robust, and is implemented with a
perturbed parameter $\lambda$ instead of the intended $\lambda_0$, one
can still obtain arbitrarily good approximations of the original
analog object without knowing the actual value of $\lambda$. However,
here we still assume that the ``perturbed'' encoder is algorithmic,
i.e., we use the same perturbed parameter value $\lambda$ at each
clock cycle. In practice, however, many circuit components are
``flaky'', that is, the associated parameters vary at each clock
cycle. We can describe such an encoder again by (\ref{alg_eqn}),
however we need to replace $(Q,F)$ by $(Q_{\lambda^n},F_{\lambda^n})$,
where $\{\lambda^n\}_1^\infty$ is the sequence of the associated
parameter vectors. With an abuse of notation, we denote such an
encoder by $\text{AE}(Q_{\lambda^n},F_{\lambda^n})$ (even though this
is clearly not an algorithmic encoder in the sense of Section \ref{AE}). Note that if
$\{\lambda^n\}_1^\infty=\lambda_0^{\N}$, i.e., $\lambda^n=\lambda_0$
for all $n$, the corresponding encoder is
$\text{AE}(Q,F,\sC(\lambda_0))$. We now say that
$\text{AE}(Q,F,\sC(\lambda_0))$ is {\em strongly robust} if there
exists $\delta>0$ such that $\text{AE}(Q_{\lambda^n},F_{\lambda^n})$
is invertible for all $\{\lambda^n\}_1^\infty$ with
$\rho_{\Lambda^\N}(\{\lambda^n\}_1^\infty,\lambda_0^\N) \le \delta$
with a common inverse. Here, $\Lambda^\N$ is the space of parameter
sequences for a given converter, and $\rho_{\Lambda^\N}$ is an
appropriate metric on $\Lambda^\N$. Similarly, we call an algorithmic
encoder {\em strongly robust in approximation} if there exists
$\delta>0$ and a family $\{D_N\}_1^\infty$ such that
$$ \delta_X(\widetilde E_N^{\text{f}},D_N) \leq C \A(E_N)$$
for all flaky encoders $\widetilde E_N^{\text{f}} = \widetilde
E_N^{\text{f}}(Q_{\lambda_n},F_{\lambda_n})$ with
$\rho_N(\{\lambda_n\}_1^N,\lambda_{0,N})<\delta$ where $\rho_N$ is the
metric on $\Lambda^N$ obtained by restricting $\rho_{\Lambda^\N}$
(assuming such a restriction makes sense), $\lambda_{0,N}$ is the
N-tuple whose components are each $\lambda_0$, and $E_N
=E_N(Q,F,\lambda_0)$.

% With an abuse of notation, we denote such encoders by
% $\text{AE}(Q,F,\sC(\lambda^n))$ (clearly this is not an algorithmic
% encoder). Note that if $\{\lambda^n\}_1^\infty=\lambda_0^{\N}$, the
% corresponding encoder is $\text{AE}(Q,F,\sC(\lambda_0))$. We now say
% that $\text{AE}(Q,F,\sC(\lambda_0))$ is {\em strongly robust} is there
% exists $\delta>0$ such that $\text{AE}(Q,F,\sC(\lambda^n))$ is
% invertible for all $\{\lambda^n\}_1^\infty\}$ with
% $\rho_{\Lambda^\N}(\{\lambda^n\}_1^\infty\},\lambda_0^\N) \le \delta$
% with a common inverse. Here, $\rho_{\Lambda^\N}$ is an appropriate
% metric in $\Lambda^\N$. Similarly, we call an algorithmic encoder {\em
%   strongly robust in approximation} by extending the above definition
% to the flaky case.

\subsubsection*{Examples of Section \ref{AE}}

Let us consider the examples of algorithmic encoders given in section
\ref{AE}. The successive approximation algorithm is a special case of
the beta encoder for $\beta = 2$ and $\tau = 1$. As we mentioned
before, there is no unique way to implement an algorithmic encoder
using a given class of components. For example, given the set of rules
set forth earlier, multiplication by $2$ could conceivably be
implemented as a replicator followed by a pair adder (though a circuit
engineer would probably not approve of this attempt). It is not our goal
here to analyze whether a given DAG model is realizable in analog
hardware, but to find out whether it is robust given its set of
parameters. 

The first order $\Sigma\Delta$ quantizer is perhaps the encoder with
the simplest DAG model; its only parametric component is the binary
quantizer, characterized by $\tau=1$.  For the beta encoder it is
straightforward to write down a DAG model that incorporates only two
parametric components: a constant multiplier and a binary quantizer.
This model is thus characterized by the vector parameter $\lambda =
(\beta, \tau)$.  The successive approximation encoder corresponds to
the special case $\lambda_0 = (2, 1)$. If the constant multiplier is
avoided via the use of a replicator and adder, as described above,
then this encoder would be characterized by $\tau = 1$, corresponding
to the quantizer threshold.

These three models ($\sd$ quantizer, beta encoder, and successive
approximation) have been analyzed in \cite{DDGV} and the following
statements hold:
\begin{enumerate}
\item The successive approximation encoder is not robust for
  $\lambda_0 =(2,1)$ (with respect to the Euclidean metric on $\Lambda
  = \R^2$). The implementation of successive approximation that avoids
  the constant multiplier as described above, and thus characterized
  by $\tau=1$ is not robust with respect to $\tau$.
\item The first order $\Sigma\Delta$ quantizer is strongly robust in
  approximation for the parameter value $\tau_0 = 1$ (and in fact for
  any other value for $\tau_0$). However, $\A(E_N) = \Theta(N^{-1})$
  for $X = [0,1]$.
\item The beta encoder is strongly robust in approximation for a range
  of values of $\tau_0$, when $\beta = \beta_0$ is fixed. Technically,
  this can be achieved by a metric which is the sum of the discrete 
metric on the first coordinate and the Euclidean metric on the second
coordinate between
  any two vectors $\lambda = (\beta, \tau)$, and $\lambda_0 =
  (\beta_0, \tau_0)$. This way we ensure that the first coordinate
remains constant on a small neighborhood of any
parameter vector. Here $\A(E_N) =
  \Theta(\beta^{-N})$ for $X = [0,1]$. This choice of the metric
however is not necessarily realistic.
\item The beta encoder is not robust with respect to the Euclidean
  metric on the parameter vector space $\Lambda=\R^2$ in which both
  $\beta$ and $\tau$ vary. In fact, this is the case even if we
  consider changes in $\beta$ only: let $E$ and $\widetilde{E}$ be the
  generators of beta encoders with parameters $(\beta,\tau)$ and
  $(\beta+\delta,\tau)$, respectively. Then $E$ and $\widetilde{E}$ do
  not have a common inverse for any $\delta \ne 0$. To see this, let
  $\{b_n\}_1^\infty=E(x)$. Then a simple calculation shows
$$ \Big |\sum_1^\infty b_n ((\beta+\delta)^{-n}-\beta^{-n})\Big | =\Omega(\frac{k}{2^{k+1}} \delta)$$
where $k=\min\{n: b_n\ne 0\}<\infty$ for any $x\ne 0$. Consequently,
$\widetilde{E}^{-1}\ne E^{-1}$.

This shows that to decode a beta-encoded bit-stream, one needs to know
the value of $\beta$ at least within the desired precision level. This
problem was addressed in \cite{DY}, where a method was proposed for
embedding the value of $\beta$ in the encoded bit-stream in such a way
that one can recover an estimate for $\beta$ (in the digital domain)
with exponential precision.  Beta encoders with the modifications of
\cite{DY} are effectively robust in approximation (the inverse of the
generator of the perturbed encoder is different from the inverse of
the intended encoder, however it can be precisely computed). Still,
even with these modifications, the corresponding encoders are
\emph{not strongly robust} with respect to the parameter $\beta$.

\item The stable $\sd$ schemes of arbitrary order $k$ that were
  designed by Daubechies and DeVore, \cite{DD}, are strongly robust in
  approximation with respect to their parameter sets. 
Also, a wide family of second-order $\sd$ schemes, as
  discussed in \cite{OY}, are strongly robust in
  approximation. On the other hand, the family of exponentially accurate
one-bit $\sd$ schemes reported in \cite{exp_decay} are not robust because
each scheme in this family employs a vector of constant multipliers
which, when perturbed arbitrarily, result in bit sequences that provide
no guarantee of even mere invertibility (using the original decoder). The 
only reconstruction accuracy guarantee is of Lipshitz type, i.e. the
error of reconstruction is controlled by a constant times the parameter distortion.

\end{enumerate}

Neither of these cases result in an algorithmic encoder with a DAG
model that is robust in its full set of parameters \emph{and}
achieves exponential accuracy.
To the best of our knowledge, our discussion in the next section 
provides the first example of an encoder that is (strongly)
robust in approximation while simultaneously achieving exponential
accuracy.

\section {The Golden Ratio Encoder} \label{GRE}

In this section, we introduce a Nyquist rate ADC, {\em the golden
  ratio encoder} (GRE), that bypasses the robustness concerns
mentioned above while still enjoying exponential accuracy in the
bit-rate. In particular, GRE is an algorithmic encoder that is
strongly robust in approximation with respect to its full set of
parameters, and its accuracy is exponential. The DAG model and block
diagram of the GRE are given in Figures \ref{DAG} and \ref{GRE_block},
respectively.

% We show that the proposed algorithm is robust with
% respect to quantizer imprecisions as well as multiplicative
% errors. Moreover, the implementation of the GRE is such that one does
% not need to generate any pre-determined real values on analog
% hardware.

% The GRE computes beta representations of real numbers
% $\beta=\phi=(1+\sqrt{5})/2$, however it does \emph{not} use the usual
% successive approximation algorithm to fullfil this task.

\subsection {The scheme} \label{GRE-1}
We start by describing the recursion underlying the GRE.  To quantize
a given real number $x\in X=[0,2]$, we set $u_0=x$ and $u_1=0$, and
run the iteration process
\begin{eqnarray}
u_{n+2}&=&u_{n+1}+u_n-b_n, \nonumber \\
b_n &=&Q(u_n,u_{n+1}). \label{rec}
\end{eqnarray}
Here, $Q:\ \R^2 \mapsto \{0,1\}$ is a quantizer to be specified later.
Note that (\ref{rec}) describes a piecewise affine discrete dynamical
system on $\mathbb{R}^2$.  More precisely, define
\begin{equation}\label{Tq}
T_Q:\begin{bmatrix} u \\ v \end{bmatrix} \mapsto \underbrace{\begin{bmatrix} 0 & 1 \\ 1 & 1 \end{bmatrix}}_{A}\begin{bmatrix} u \\ v \end{bmatrix} - Q(u,v) \begin{bmatrix} 0 \\ 1 \end{bmatrix}.
\end{equation}
Then we can rewrite (\ref{rec}) as 
\begin{equation} \label{rec2}
\begin{bmatrix} u_{n+1} \\ u_{n+2} \end{bmatrix} = T_Q\begin{bmatrix*}[l] u_{n} \\ u_{n+1}\end{bmatrix*}.
\end{equation}
Now, let $X=[0,2)$, $\U\subset \R^2$, and suppose $Q$ is a quantizer
on $X \times \U$. The formulation in (\ref{rec2}) shows that GRE is an
algorithmic encoder, $\text{AE}(Q,F_{\text{GRE}})$, where
$F_{\text{GRE}}(x,\mathbf{u}):=A\mathbf{u}-Q(\mathbf{u})$ for
$\mathbf{u}\in \U$.  Next, we show that GRE is invertible by
establishing that, if $Q$ is chosen appropriately, the sequence
$b=(b_n)$ obtained via (\ref{rec}) gives a beta representation of $x$
with $\beta=\phi=(1+\sqrt{5})/2$.

\subsection {Approximation Error and Accuracy} \label{GRE-2}
\begin{prop}\label{prop_rec}
  Let $x\in [0,2)$ and suppose $b_n$ are generated via {\em(\ref{rec})}
  with $u_0=x$ and $u_1=0$. Then
$$x=\sum_{n=0}^\infty b_n \phi^{-n}$$
if and only if the state sequence $(u_n)_0^\infty$ is
bounded. Here $\phi$ is the golden mean.
\end{prop}
\begin{IEEEproof}
Note that 
\begin{eqnarray}
\sum_{n=0}^{N-1} b_{n} \phi^{-n} &=& \sum_{n=0}^{N-1} (u_n+u_{n+1}-u_{n+2}) \phi^{-n} \nonumber \\
% &=& \sum_{n=0}^{N-1} u_n \phi^{-n} + \sum_{n=1}^{N} u_n \phi^{-n+1} - \sum_{n=2}^{N+1} u_n \phi^{-n+2} \nonumber \\
&=& \sum_{n=2}^{N-1} u_n \phi^{-n}(1+\phi-\phi^2) + u_1 + u_{N}\phi^{-N+1} + u_0 + u_1 \phi^{-1}-u_{N}\phi^{-N+2}-u_{N+1}\phi^{-N+1} \nonumber \\
&=& x+\phi^{-N+1}((1-\phi)u_{N}-u_{N+1}), \nonumber \\
&=& x-\phi^{-N}(u_N + \phi\, u_{N+1}) \label{N-term-equiv}.
\end{eqnarray}
where the third equality follows from $1+\phi-\phi^2=0$, and the last
equality is obtained by setting $u_0=x$ and $u_1=0$. Defining the
$N$-term approximation error to be
$$
e_N(x) := x - \sum_{n=0}^{N-1} b_{n} \phi^{-n}= \phi^{-N}(u_N + \phi\, u_{N+1})$$
it follows that 
$$
\lim_{N\to \infty} e_N(x) = 0,
$$
provided 
\begin{equation} \label{subc1}
\lim_{n\to \infty} u_n \phi^{-n}=0.
\end{equation}

Clearly, (\ref{subc1}) is satisfied if there is a constant $C$,
independent of $n$, such that $|u_n|\le C$, . Conversely, suppose
(\ref{subc1}) holds, and assume that $(u_n)$ is unbounded. Let N be
the smallest integer for which $|u_N|>C'\phi^3$ for some
$C'>1$. Without loss of generality, assume $u_N>0$ (the argument
below, with simple modifications, applies if $u_N<0$). Then, using
(\ref{rec}) repeatedly, one can show that
$$u_{N+k}>C'\phi^3 f_{k-1} - f_{k+2}+1$$
where $f_k$ is the $k$th Fibonacci number. Finally, using 
$$f_k=\frac{\phi^k -(1-\phi)^k}{\sqrt{5}}
$$
(which is known as Binet's formula, e.g., see \cite{V}) we conclude
that $u_{N+k}\ge C''\phi^k$ for every positive integer $k$, showing
that (\ref{subc1}) does not hold if the sequence $(u_n)$ is not
bounded.

\end{IEEEproof}

% \begin{remark} One can show that the state sequence $(u_n)$ obtained by (\ref{}) is either bounded, or $u_n=\Theta(\phi^n)$. Consequently,   

Note that the proof of Proposition \ref{prop_rec} also shows that the
$N$-term approximation error $e_N(x)$ decays exponentially in $N$ if
and only if the state sequence $(u_n)$, obtained when encoding $x$,
remains bounded by a constant (which may depend on $x$). We will say
that a GRE is {\em stable} on $X$ if the constant $C$ in Proposition
\ref{prop_rec} is independent of $x\in X$, i.e., if the state
sequences $u$ with $u_0=x$ and $u_1=0$ are bounded by a constant
uniformly in $x$. In this case, the following proposition holds.

\begin{prop} \label{exp_acc}
  Let $E^{\text{GRE}}$ be the generator of the GRE, described by
  {\em(\ref{rec})}. If the GRE is stable on $X$, it is exponentially
  accurate on $X$. In particular, $\A(E^{GRE}_N)=\Theta(\phi^{-N})$.

% Suppose $Q$ in (\ref{rec}) is such that the state sequence with $u_0=x$ and $u_1=0$ satisfies $\|u\|_{\infty} \le C$. Then, 
% $$
% |e_N(x)| = \left |x-\sum_{n=0}^{N-1} b_n \phi^{-n} \right | \le C \phi^{-N+2}.
% $$ 
\end{prop}
Next, we investigate quantizers $Q$ that generate stable encoders.

\subsection{Stability and robustness with respect to imperfect quantizers} \label{GRE-3}
To establish stability, we will show that for several choices of the quantizer
$Q$, there exist bounded positively invariant sets $R_Q$ such that 
$T_Q(R_Q)\subseteq R_Q$. We will frequently use the basic 1-bit quantizer,
$$
q_1(u)=
\begin{cases}
0, & \text{if}\ u < 1, \\
1, & \text{if}\ u \geq 1.
\end{cases}
$$
Most practical quantizers are implemented using arithmetic
operations and $q_1$. One class that we will consider is given by
\begin{equation} \label{Qalpha}
Q_\alpha(u,v) := q_1(u+\alpha v) = 
\begin{cases}
0, & \text{if}\ u+\alpha v < 1, \\
1, & \text{if}\ u+\alpha v \geq 1.
\end{cases}
\end{equation}
Note that in the DAG model of GRE, the circuit components that
implement $Q_\alpha$ incorporate a parameter vector
$\lambda=(\tau,\alpha)=(1,\alpha)$. Here, $\tau$ is the threshold of the
1-bit basic quantizer, and $\alpha$ is the gain factor of the 
multiplier that maps $v$ to $\alpha v$. 
% introduces two components in the DAG model of GRE, shown in Figure
% \ref{DAG}: $\tau=1$ for the threshold $\tau$ of the 1-bit basic
% quantizer, and the multiplier gain $\alpha$.
One of our main goals in this paper is to prove that GRE, with the
implementation depicted in Figure \ref{GRE_block}, is strongly robust
in approximation with respect to its full set of parameters. That is,
we shall allow the parameter values to change at each clock cycle
(within some margin). Such changes in parameter $\tau$ can be
incorporated to the recursion (\ref{rec}) by allowing the quantizer
$Q_\alpha$ to be flaky. More precisely, for $\nu_1 \leq \nu_2$, let
$q^{\nu_1,\nu_2}$ be the flaky version of $q_\tau$ defined by
$$
q^{\nu_1,\nu_2}(u) :=
\begin{cases}
0, & \text{if}\ u < \nu_1, \\
1, & \text{if}\ u \geq \nu_2, \\
0\ \text{or}\ 1, & \text{if}\ \nu_1 \le u < \nu_2.
\end{cases}
$$
We shall denote by $Q^{\nu_1,\nu_2}_\alpha$ the 
{\em flaky} version of $Q_\alpha$,
which is now
$$Q^{\nu_1,\nu_2}_\alpha(u,v) := q^{\nu_1,\nu_2}(u+\alpha v).$$
Note that (\ref{rec}), implemented with $Q=Q^{\nu_1,\nu_2}_\alpha$,
does not generate an algorithmic encoder. At each clock cycle, the
action of $Q^{\nu_1,\nu_2}_\alpha$ is identical to the action of
$Q^{\tau_n,\tau_n}_\alpha(u,v) := q_{\tau_n}(u+\alpha v)$ for some
$\tau_n \in [\nu_1,\nu_2)$. In this case, using the notation
introduced before, (\ref{rec}) generates
$\text{AE}(F^{\tau_n}_{\text{GRE}},Q^{\tau_n,\tau_n}_\alpha)$. We will refer
to this encoder family as \emph{GRE with flaky quantizer}.

\subsubsection{A stable GRE with no multipliers: the case $\alpha=1$}
We now set $\alpha=1$ in (\ref{Qalpha}) and show that the GRE
implemented with $Q_1$ is stable, and thus generates an encoder family
with exponential accuracy (by Proposition \ref{exp_acc}). Note that in
this case, the recursion relation (\ref{rec}) does not employ any
multipliers (with gains different from unity). In other words, the
associated DAG model does not contain any ``constant multiplier''
component.

\begin{prop} \label{gre_stab_id} Consider $T_{Q_1}$, defined as in
  {\em(\ref{Tq})}. Then $R_{Q_1}:=[0,1]^2$ satisfies
$$
T_{Q_1}(R_{Q_1})= R_{Q_1}.
$$
\end{prop}
\begin{IEEEproof}
  By induction. It is easier to see this on the equivalent recursion
  (\ref{rec}). Suppose $(u_n,u_{n+1}) \in R_{Q_1}$, i.e., $u_n$ and
  $u_{n+1}$ are in $[0,1]$. Then
  $u_{n+2}=u_n+u_{n+1}-q_1(u_n+u_{n+1})$ is in $[0,1]$, which
  concludes the proof.
\end{IEEEproof} 

It follows from the above proposition that the GRE implemented with
$Q_1$ is stable whenever the initial state $(x,0) \in R_{Q_1}$, i.e.,
$x\in [0,1]$. In fact, one can make a stronger statement because a
longer chunk of the positive real axis is in the basin of attraction
of the map $T_{Q_1}$.

\begin{prop} \label{prop_gre_id} The GRE implemented with $Q_1$ is
  stable on $[0,1+\phi)$, where $\phi$ is the golden mean. More
  precisely, for any $x \in [0,1+\phi)$, there exists a positive
  integer $N_x$ such that $ u_n \in [0,1]$ for all $n\ge N_x$.
\end{prop}

\begin{cor} \label{cor_gre_id} Let $x \in [0,1+\phi)$, set $u_0=x$,
  $u_1=0$, and generate the bit sequence $(b_n)$ by running the
  recursion {\em(\ref{rec})} with $Q=Q_1$. Then, for $N\ge N_x$,
$$ 
|x-\sum_{n=0}^{N-1} b_n \phi^{-n}| \le \phi^{-N+2}.
$$
One can choose $N_x$ uniformly in $x$ in any closed subset of
$[0,1+\phi)$. In particular, $N_x=0$ for all $x\in [0,1]$.
\end{cor}

\begin{figure}[t] \textmd{}
  % \centerline{\includegraphics[width=3in]{figQ1_0_01} \hskip
%     1cm \includegraphics[width=3in]{figQ1_0_1}}
\centerline{\includegraphics[width=3in]{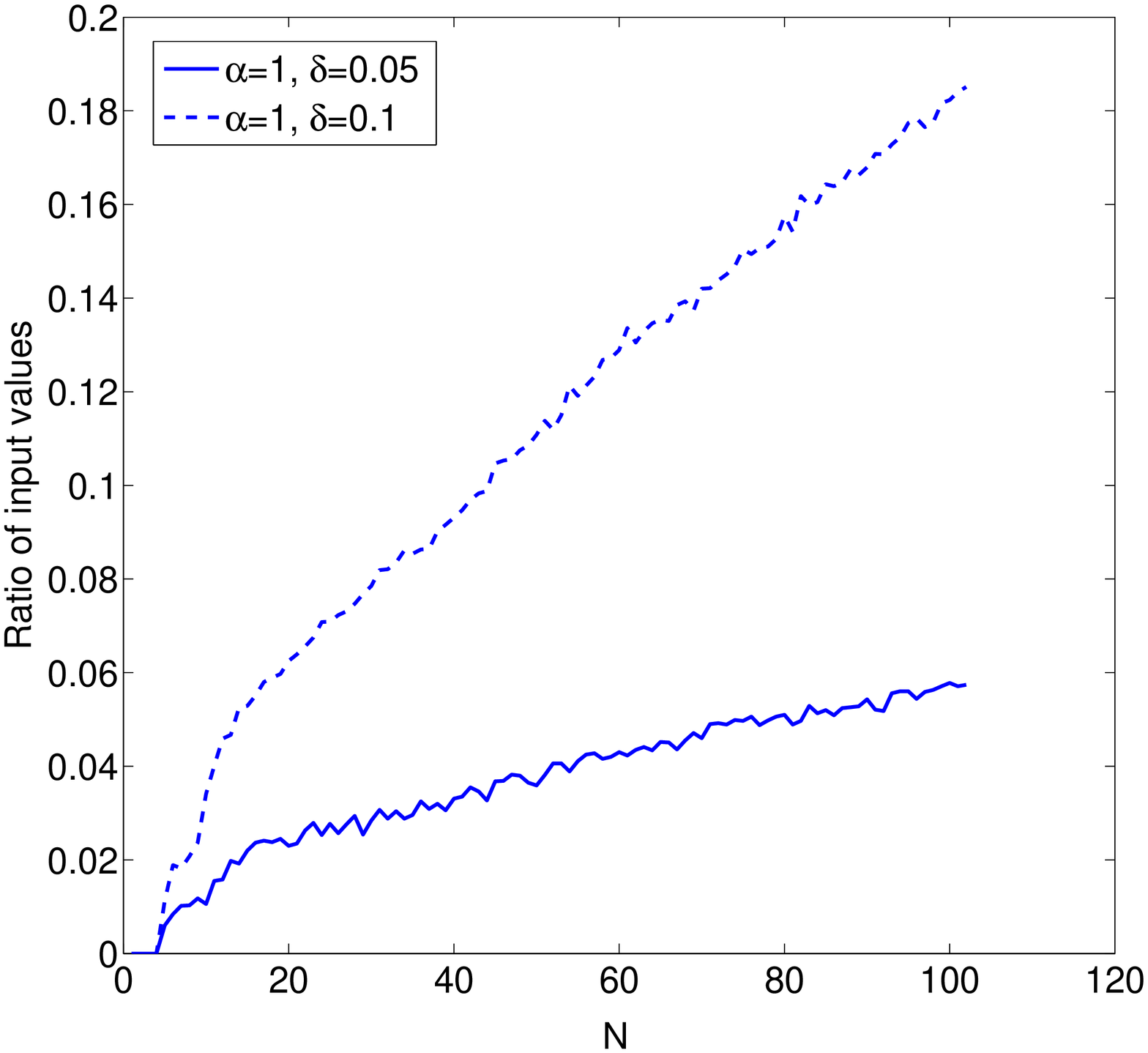}}
\caption{
\label{figQ1}
We choose 10,000 values for $x$, uniformly distributed in $[0,1]$ and
run the recursion (\ref{rec}) with $Q=Q_1^{\nu_1,\nu_2}$ with $\nu_i
\in (1-\delta, 1+\delta)$. The graphs show the ratio of the input
values for which $|u_N|>1$ vs. $N$ for $\delta=0.05$ (solid) and  
$\delta=0.1$ (dashed).}
\end{figure}

\noindent
{\bf Remarks.} \begin{enumerate}
\item The proofs of Proposition \ref{prop_gre_id} and Corollary
  \ref{cor_gre_id} follow trivially from Proposition \ref{gre_stab_id}
  when $x \in [0,1]$.  It is also easy to see that $N_x=1$ for $x\in
  (1,2]$, i.e., after one iteration the state variables $u_1$ and
  $u_2$ are both in $[0,1]$. Furthermore, it can be shown that
$$ 0<x<1+\frac{f_{N+2}}{f_{N+1}}-\frac{1}{f_{N+1}} \ \ \Rightarrow \ \ N_x \le N,
$$
where $f_n$ denotes the $n$th Fibonacci number. We do not include the
proof of this last statement here, as the argument is somewhat long,
and the result is not crucial from the viewpoint of this paper.

\item In this case, i.e., when $\alpha=1$, the encoder is {\em not
    robust} with respect to quantizer imperfections. More precisely,
  if we replace $Q_1$ with $Q_1^{\nu_1,\nu_2}$, with $\nu_1,\nu_2 \in
  (1-\delta,1+\delta)$, then $|u_n|$ can grow regardless of how small
  $\delta>0$ is. This is a result of the mixing properties of the
  piecewise affine map associated with GRE. In particular, one can
  show that $ [0,1] \times \{0\} \subset \displaystyle \cup_n S_n$,
  where $S_n \subset [0,1]^2$ is the set of points whose $n$th forward
  image is outside the unit square. Figure \ref{figQ1} shows the
  fraction of 10,000 randomly chosen $x$-values for which $|u_N|>1$ as
  a function of $N$. In fact, suppose $u_0=x$ with $x\in [0,1]$ and
  $u_1=0$. Then, the probability that $u_N$ is outside of $[0,1]$ is
  $O(N\delta^2+\phi^{-N})$, which is superior to the case with PCM
  where the corresponding probability scales like $N\delta$. This
  observation suggests that ``GRE with no multipliers'', with its
  simply-implementable nature, could still be useful in applications
  where high fidelity is not required. We shall discuss this in more
  detail elsewhere.
\end{enumerate}

\subsubsection{Other values of $\alpha$: stable and robust GRE}

% In general, we have that for a perfect (i.e., not flaky)
% quantizer $Q$, if the map $T_Q$ is stable then it possesses tiling invariant
% sets. The proof of this result will follow from arguments that are 
% similar to the maps derived from sigma-delta modulation. We will come to this
% later. Some examples are shown in Figure \ref{tiles}.

% \begin{figure}[ht]
% \begin{center}
% \includegraphics[width=2in]{tile1} 
% \includegraphics[width=2in]{tile2} \\
% \includegraphics[width=2in]{tile3} 
% \includegraphics[width=2in]{tile4}
% \end{center}
% \caption{
% \label{tiles}
% Various tiling invariant sets (in blue) for perfect quantizers
% and their translates by $(0,1)$ and $(1,0)$ in green and red, respectively. Thresholding line is given by the black line.}
% \end{figure}

%\subsection{Stability under additive noise and the effect of arithmetic errors}
In the previous subsection, we saw that GRE, implemented with
$Q_\alpha$, $\alpha=1$, is stable on $[0,1+\phi)$, and thus enjoys
exponential accuracy; unfortunately, the resulting algorithmic encoder
is not robust. In this subsection, we show that there is a wide
parameter range for $\nu_1,\nu_2$ and $\alpha$ for which the map $T_Q$
with $Q = Q_\alpha^{\nu_1,\nu_2}$ has positively invariant sets $R$
that do not depend on the particular values of $\nu_i$ and
$\alpha$. Using such a result, we then conclude that, with the
appropriate choice of parameters, the associated GRE is strongly
robust in approximation with respect to $\tau$, the quantizer
threshold, and $\alpha$, the multiplier needed to implement
$Q_\alpha$. We also show that the invariant sets $R$ can be
constructed to have the additional property that for a small value of
$\mu > 0$ one has $T_Q(R) + B_\mu(0) \subset R$, where $B_\mu(0)$
denotes the open ball around 0 with radius $\mu$. In this case, $R$
depends on $\mu$. Consequently, even if the image of any state $(u,v)$
is perturbed within a radius of $\mu$, it still remains within
$R$. Hence we also achieve stability under small additive noise or
arithmetic errors.

\begin{lm} Let $0\le \mu \le (2\phi^2\sqrt{\phi+2})^{-1}\approx
  0.1004$. There is a set $R=R(\mu)$, explicitly given in \eqref{invseteqn},
  and a wide range for parameters $\nu_1,\nu_2$, and $\alpha$ such
  that $T_{Q_\alpha^{\nu_1,\nu_2}}(R) \subseteq R+B_\mu(0)$.
\end{lm}

\begin{IEEEproof} Our proof is constructive. In particular, for a given
  $\mu$, we obtain a parametrization for $R$, which turns out
  to be a rectangular set. The corresponding ranges for $\nu_1,\nu_2$,
  and $\alpha$ are also obtained implicitly below. We will give
  explicit ranges for these parameters later in the text.  

  Our construction of the set $R$, which only depends on $\mu$, is
  best explained with a figure. Consider the two rectangles
  $A_1B_1C_1D_1$ and $A_2B_2C_2D_2$ in Figure \ref{invariant_set}.
  These rectangles are designed to be such that their respective
  images under the linear map $T_1$, and the affine map $T_2$, defined
  by
\begin{equation}
T_1:\begin{bmatrix} u \\ v \end{bmatrix} \mapsto \begin{bmatrix} 0 & 1 \\ 1 & 1 \end{bmatrix},
\;\;\;\;\mbox{and} \;\;\;
T_2:\begin{bmatrix} u \\ v \end{bmatrix} \mapsto \begin{bmatrix} 0 & 1 \\ 1 & 1 \end{bmatrix}
\begin{bmatrix} u \\ v \end{bmatrix} - \begin{bmatrix} 0 \\ 1 \end{bmatrix},
\end{equation}
are the same, i.e., $T_1(A_1B_1C_1D_1)=ABCD=T_2(A_2B_2C_2D_2)$.

\begin{figure}[ht]
\centerline{\includegraphics[width=6in]{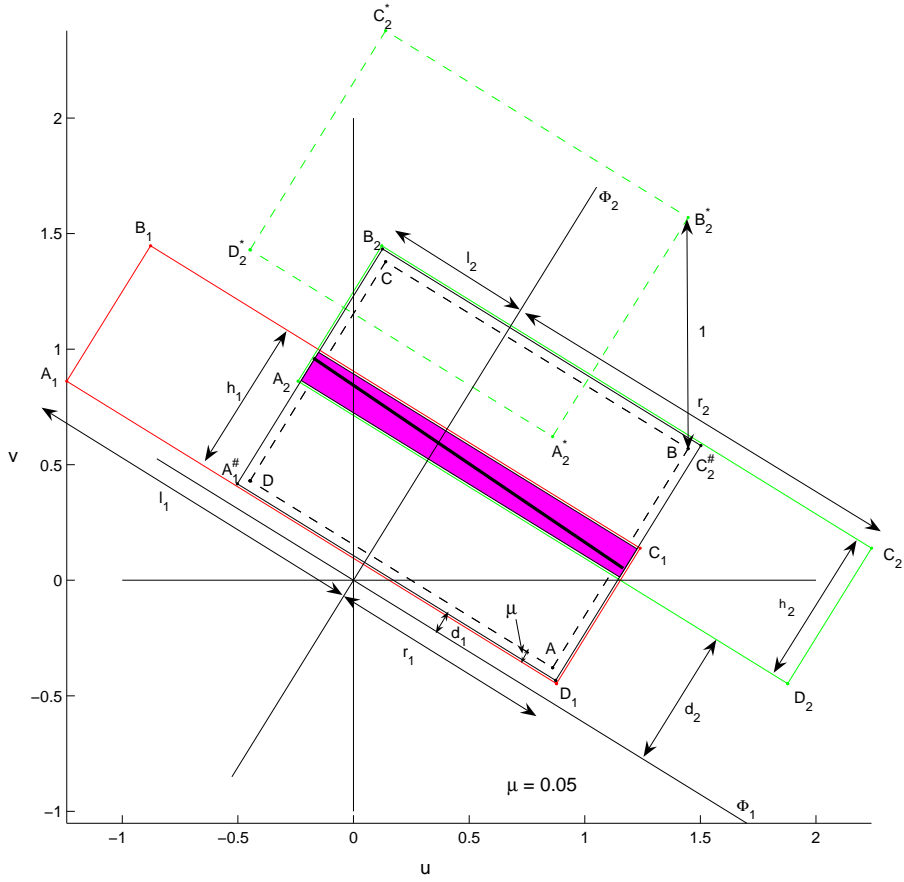}}
%\centerline{\includegraphics[width=4.5in]{invariant_figure}}
\caption{
\label{invariant_set}
A positively invariant set and the demonstration of its robustness 
with respect to additive noise and imperfect quantization.}
\end{figure}

The rectangle $ABCD$ is such that its $\mu$-neighborhood is contained
within the union of $A_1B_1C_1D_1$ and $A_2B_2C_2D_2$. This guards
against additive noise. The fact that the rectangles $A_1B_1C_1D_1$
and $A_2B_2C_2D_2$ overlap (the shaded region) allows for the use of
a flaky quantizer. Call this region $F$.  As long as the region in
which the quantizer operates in the flaky mode is a subset of $F$, and
$T_Q = T_1$ on $A_1B_1C_1D_1 \setminus F$ and $T_Q = T_2$ on
$A_2B_2C_2D_2 \setminus F$, it follows that $T_Q(A_1B_1C_1D_1 \cup
A_2B_2C_2D_2) \subset ABCD$.  It is then most convenient to choose
$R=R(\mu) = A^\#_1B_2C^\#_2D_1$ and we clearly have $T_Q(R)+B_\mu(0)
\subset R$. Note that if $Q=Q_\alpha^{\nu_1,\nu_2}$, any choice
$\nu_1,\nu_2$, and $\alpha$ for which the graph of $v=-\frac{1}{\alpha} u +
\frac{\nu}{\alpha}$ remains inside the shaded region $F$ for $\nu_1\le \nu \le
\nu_2$ will ensure $T_Q(R)\subseteq R+B_\mu(0)$.

Next, we check the existence of at least one solution to this
setup. This can be done easily in terms the parameters defined in the
figure.  First note that the linear map $T_1$ has the eigenvalues
$-1/\phi$ and $\phi$ with corresponding (normalized) eigenvectors
$\Phi_1 = \frac{1}{\sqrt{\phi+2}}(\phi,\;-1)$ and $\Phi_2 =
\frac{1}{\sqrt{\phi+2}} (1,\; \phi)$.  Hence $T_1$ acts as an
expansion by a factor of $\phi$ along $\Phi_2$, and reflection
followed by contraction by a factor of $\phi$ along $\Phi_1$. $T_2$ is
the same as $T_1$ followed by a vertical translation of $-1$.  It
follows after some straightforward algebraic calculations that the
mapping relations described above imply
\begin{eqnarray}
r_1 &=&  \frac{\phi}{\sqrt{\phi+2}} + \phi^2 \mu \nonumber \\
l_1 &=&  \frac{\phi^2}{\sqrt{\phi+2}} + \phi^2 \mu \nonumber \\
r_2 &=&  \frac{2 \phi}{\sqrt{\phi+2}} + \phi^2 \mu \nonumber\\
l_2 &=&  \frac{1}{\sqrt{\phi+2}} + \phi^2 \mu \nonumber \\
h_1\,=\,h_2 &=& \frac{\phi}{\sqrt{\phi+2}} - 2\phi \mu \nonumber \\
d_1 &=& \phi \mu \nonumber \\
d_2 &=& \frac{1}{\sqrt{\phi+2}} + \phi \mu. \nonumber
\end{eqnarray}
Consequently, the positively invariant set $R$ is the set of all points inside the rectangle $ A^\#_1B_2C^\#_2D_1$ where
\begin{eqnarray}
A^\#_1 &=& -l_2 \Phi_1 +  d_1 \Phi_2  \nonumber \\ 
B_2 &=& -l_2 \Phi_1 + (d_2+h_2) \Phi_2 \nonumber \\
C^\#_2 &=& r_1 \Phi_1 + (d_2+h_2) \Phi_2 \nonumber \\
D_1 &=& r_1 \Phi_1 +  d_1 \Phi_2 \label{invseteqn}
\end{eqnarray}
Note that $R$ depends only on $\mu$. Moreover, the existence of the
overlapping region $F$ is equivalent to the condition $d_1+h_1 > d_2$
which turns out to be equivalent to 
$$\mu < \frac{1}{2\phi^2 \sqrt{\phi+2}} \approx 0.1004.$$
\end{IEEEproof}

\subsubsection*{Flaky quantizers, linear thresholds}
Next we consider the case $Q=Q_\alpha^{\nu_1,\nu_2}$ and specify ranges for $\nu_1,\nu_2$, and $\alpha$ such that $T_Q(R(\mu)) \subseteq R(\mu)+B_\mu(0)$. 

\begin{prop} \label{ranges}
Let $0<\mu < \frac{1}{2\phi^2 \sqrt{\phi+2}}$
and $\alpha$ such that 
\begin{equation} \label{alpharange}
\alpha_{\text{min}}(\mu):=1+2\mu \phi \sqrt{\phi+2} \le \alpha \le 3-\frac{10\mu\phi\sqrt{\phi+2}}{1+4\mu\sqrt{\phi+2}}=:\alpha_{\text{max}}(\mu)
\end{equation}
be fixed. Define
\begin{eqnarray}
\nu_{min}(\alpha,\mu)&:=&
\begin{cases}
1+\mu \phi \sqrt{\phi+2}, & \alpha \le \phi, \\
\alpha\left(\frac{\phi+1}{\phi+2}+\frac{2\mu\phi^2}{\sqrt{\phi+2}}\right) 
+\left(\frac{1-\phi}{\phi+2}-\frac{\mu\phi^2}{\sqrt{\phi+2}}\right), & \alpha>\phi,
\end{cases} \\
\nu_{max}(\alpha,\mu)&:=&
\begin{cases}
\alpha-\mu\phi\sqrt{\phi+2}, & \alpha \le \phi, \\
\alpha\left(\frac{1}{\phi+2}-\frac{2\mu\phi^2}{\sqrt{\phi+2}}\right) 
+\left(\frac{1+2\phi}{\phi+2}+\frac{\mu\phi^2}{\sqrt{\phi+2}}\right), & \alpha>\phi.
\end{cases}
\end{eqnarray}
If $\nu_{min}(\alpha,\mu_0) \le \nu_1 \le \nu_2 \le
\nu_{max}(\alpha,\mu_0)$, then $T_{Q_\alpha^{\nu_1,\nu_2}}(R(\mu))
\subseteq R(\mu)+B_\mu(0)$ for every $\mu \in [0,\mu_0]$.
\end{prop}
\begin{figure}[t]
\centerline{\includegraphics[width=3.5in]{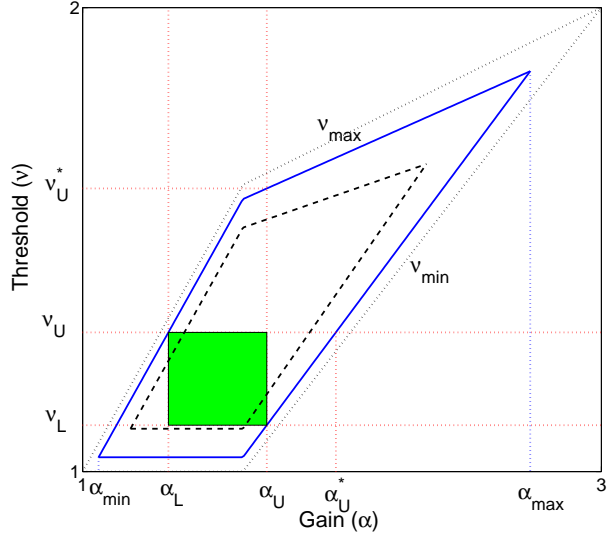}}
\caption{
\label{nu_alp}
We plot $\nu_{min}(\alpha,\mu)$ and $\nu_{min}(\alpha,\mu)$ with
$\mu=0.03$ (dashed), $\mu=0.01$ (solid), and $\mu=0$ (dotted). If
$\{\alpha\} \times [\nu_1,\nu_2]$ remains in the shaded region, then
$T_{Q^{\nu_1,\nu_2}_\alpha}(R(\mu)) \subseteq R(\mu) + B_\mu(0)$ for
$\mu=0.01$.}
\end{figure}

\noindent
{\bf Remarks.}
\begin{enumerate}
\item The dark line segment depicted in Figure \ref{invariant_set}
  within the overlapping region $F$, refers to a hypothetical
  quantizer threshold that is allowed to vary within $F$. Proposition
  \ref{ranges} essentially determines the vertical axis intercepts of
  lines with a given slope, $-1/\alpha$, the corresponding segments of
  which remain in the overlapping region $F$. The proof is
  straightforward but tedious, and will be omitted.
\item In Figure~\ref{nu_alp} we show $\nu_{min}(\alpha,\mu)$ and
  $\nu_{max}(\alpha,\mu)$ for $\mu=0,0.01,0.03$. Note that $\nu_{min}$
  and $\nu_{max}$ are both increasing in $\alpha$.  Moreover, for
  $\alpha$ in the range shown in \eqref{alpharange}, we have
  $\nu_{min}(\alpha,\mu) \le \nu_{max}(\alpha,\mu)$ with
  $\nu_{max}(\alpha,\mu)=\nu_{min}(\alpha,\mu)$ at the two endpoints,
  $\alpha_{min}(\mu)$ and $\alpha_{min}(\mu)$. Hence, $\nu_{min}$ and
  $\nu_{max}$ enclose a bounded region, say $G(\mu)$. If $\{\alpha\}
  \times [\nu_1,\nu_2]$ is in $G(\mu)$, then $T_Q(R(\mu)) \subseteq
  R(\mu)+B_\mu(0)$ for $Q=Q^{\nu_1,\nu_2}_\alpha$.

\item For any $\alpha_L \in (\alpha_{min}(\mu),\alpha_{max}(\mu))$, note that 
$\nu_{min}$ is invertible at $\alpha_L$ and set 
$$ \alpha_U^*:=\nu_{min}^{-1}(\nu_{max}(\alpha_L)).
$$
Then, for any $\alpha_U \in (\alpha_L,\alpha_U^*)$, we have 
$$ (\alpha_L,\alpha_U) \times (\nu_1,\nu_2) \in G(\mu)
$$
provided 
$$ \nu_L:=\nu_{min}(\alpha_U)\le \nu_1 < \nu_2 \le \nu_{max}(\alpha_L)=:\nu_U.
$$
Note also that for any $(\alpha_1,\alpha_2) \subset (\alpha_L,\alpha_U^*)$, we have $\nu_{min}(\alpha_2) < \nu_{max}(\alpha_1)$. Thus,
$$ (\alpha_1,\alpha_2)\times (\nu_1,\nu_2) \subset G(\mu)$$
if $\nu_{min}(\alpha_2) \le \nu_1 < \nu_2 \le \nu_{max}(\alpha_1)$.
Consequently, we observe that
$$T_{Q^{\nu_1,\nu_2}_\alpha}(R(\mu))\subseteq R(\mu)+B(\mu)$$ 
for any $\alpha \in (\alpha_1,\alpha_2)$ and 
$[\nu_1,\nu_2]\subset [\nu_{min}(\alpha_2),\nu_{max}(\alpha_1)]$. 
\item We can also determine the allowed range for $\alpha$, given the
  range for $\nu_1,\nu_2$, by reversing the argument above. The extreme
  values of $\nu_{min}$ and $\nu_{max}$ are
  \begin{eqnarray}
    \nu_{min}(\alpha_{min})&=&1+\mu\phi\sqrt{\phi+2}, \\
    \nu_{min}(\alpha_{max}) &=& \frac{6\phi+2+3\mu(4\phi+3)\sqrt{\phi+2}}{(\phi+4
\mu \phi^2 \sqrt{\phi+2})(\phi+2)}.
\end{eqnarray}
For any $\nu_L$ in the open interval between these extreme values, set 
$$\nu_U^*:=\nu_{max}(\nu_{min}^{-1}(\nu_L)).
$$
Then, for any $\nu_U \in (\nu_L,\nu_U^*)$,
$(\alpha_L,\alpha_U) \times (\nu_L,\nu_U)$ is in the allowed region $G(\mu)$ where 
\begin{eqnarray}
\alpha_L:=\nu_{max}^{-1}(\nu_U)  \\
\alpha_U:=\nu_{min}^{-1}(\nu_L).
\end{eqnarray}
This is shown in Figure \ref{nu_alp}. Consequently, we observe that
GRE implemented with $Q=Q^{\nu_1,\nu_2}_\alpha$ remains stable for any
$\nu_1,\nu_2$ and $\alpha$ provided $[\nu_1,\nu_2] \subset
(\nu_L,\nu_U)$ and $\alpha \in (\alpha_L,\alpha_U)$. 

\item For the case $\mu=0$, the expressions above simplify
  significantly. In \eqref{alpharange}, $\alpha_{min}(0)=1$ and
  $\alpha_{max}(0)=3$. Consequently, the extreme values $\nu_{min}$
  and $\nu_{max}$ are 1 and 2, respectively. One can repeat the
  calculations above to derive the allowed ranges for the parameters
  to vary. Observe that $\mu=0$ gives the widest range for the
  parameters. This can be seen in Figure~\ref{nu_alp}.
\end{enumerate}

We now go back to the GRE and state the implications of the results
obtained in this subsection. The recursion relations \eqref{rec} that
define GRE assume perfect arithmetic. We modify \eqref{rec} to allow
arithmetic imperfections, e.g., additive noise, as follows
\begin{eqnarray}
u_{n+2}&=&u_{n+1}+u_n-b_n+\epsilon_n, \nonumber \\
b_n &=&Q(u_n,u_{n+1}), \label{rec-n}
\end{eqnarray}
and conclude this section with the following stability theorem.
\begin{thm} \label{stab_thm} Let $\mu \in
  [0,\frac{1}{2\phi^2\sqrt{\phi+2}})$, $\alpha_{min}(\mu)$ and
  $\alpha_{max}(\mu)$ as in \eqref{alpharange}. For every \\ $\alpha
  \in (\alpha_{min}(\mu),\alpha_{max}(\mu))$, there exists
  $\nu_1,\nu_2$, and $\eta>0$ such that the encoder described by
  \eqref{rec-n} with $Q=Q_{\alpha'}^{\nu_1,\nu_2}$ is stable provided
  $|\alpha'-\alpha| < \eta$ and $|\epsilon_n| < \mu$.
\end{thm}
\begin{IEEEproof}
  This immediately follows from our remarks above. In particular,
  given $\alpha$, choose $\alpha_L<\alpha$ such that the corresponding
  $\alpha_U^*=\nu_{min}^{-1}(\nu_{max}(\alpha_L)) >\alpha$. By
  monotonicity of both $\nu_{min}$ and $\nu_{max}$, any $\alpha_L \in
  (\nu_{max}^{-1}(\nu_{min}(\alpha)),\alpha)$ will do.
  Next, choose some $\alpha_U \in (\alpha,\alpha_U^*)$, and
  set $\eta:=\min\{\alpha-\alpha_L,\alpha_U-\alpha \}$. The statement
  of the theorem now holds with $\nu_1=\nu_{min}(\alpha_U)$ and
  $\nu_2=\nu_{max}(\alpha_L)$.
\end{IEEEproof}

When $\mu=0$, i.e., when we assume the recursion relations \eqref{rec}
are implemented without an additive error, Theorem \ref{stab_thm}
implies that GRE is strongly robust in approximation with respect to
the parameter $\alpha,\nu_1$, and $\nu_2$. More precisely, the
following corollary holds.

\begin{cor} \label{cor_main} Let $x\in [0,1]$, and $\alpha \in
  (1,3)$. There exist $\eta>0$, and $\nu_1,\nu_2$ such that $b_n$,
  generated via \eqref{rec} with $u_0=x$, $u_1=0$, and
  $Q=Q_{\alpha'}^{\nu_1,\nu_2}$, approximate $x$ exponentially
  accurately whenever $|\alpha'-\alpha| < \eta$. In particular, the
  $N$-term approximation error $e_N(x)$ satisfies
$$e_N(x)=x-\sum_{n=0}^{N-1} b_n \phi^{-n} \le C \phi^{-N}
$$
where $C=1+\phi$.
\end{cor}
\begin{IEEEproof} The claim follows from Proposition \ref{prop_rec} and Theorem
  \ref{stab_thm}: Given $\alpha$, set $\mu=0$, and choose
  $\nu_1,\nu_2$, and $\eta$ as in Theorem \ref{stab_thm}. As the set
  $R(0)$ contains $[0,1] \times \{0\}$, such a choice ensures that the
  corresponding GRE is stable on $[0,1]$. Using Proposition \ref{prop_rec}, we
  conclude that $e_N(x) \le \phi^{-N}(u_N+\phi u_{N+1})$. Finally, as
  $(u_N,u_{N+1}) \in R(0)$, 
  $$ u_N+\phi u_{N+1} \le \max_{(u,v)\in
    R(0)} u+\phi v=1+\phi.$$
\end{IEEEproof}

% Consider the angle $\psi$ that this line makes 
% with the direction $-\Phi_1$, measured in the clockwise direction.
% It follows that
% $$ -\tau \leq \tan \psi  \leq \tau,$$
% where 
% $ \tau := \tau(\mu) := \frac{d_1 + h_1 - d_2}{r_1 + l_2}.$

% If the flaky quantizer $Q = Q_\alpha^{\nu_1,\nu_2}$ is used, then we easily 
% obtain the chain of inequalities
% $$ 
% \frac{\phi^{-1} - \tau}{1 + \phi^{-1}\tau} \leq
% \frac{\phi^{-1} - \tan \psi}{1 + \phi^{-1}\tan \psi} \leq
% \alpha^{-1} \leq \frac{\phi^{-1} + \tan \psi}{1- \phi^{-1}\tan \psi}
% \leq \frac{\phi^{-1} +\tau}{1-\phi^{-1}\tau}
% $$
% It is easy to see that the largest possible value of $\tau$, and hence
% the widest range for $\psi$, occurs
% when $\mu = 0$. This gives us the maximum attainable 
% $\tau = \phi^{-3}$. Plugging this value in the bounds 
% and using the relation $\phi^2 = \phi+1$ repeatedly results in the global 
% bound for the parameter $\alpha$:
% $ 1 \leq \alpha \leq 3$.

% For $\mu > 0$, using the corresponding expression for $\tau (\mu)$ 
% implies that for each value of $\alpha$ satisfying $1 < \alpha < 3$,
% there exist admissible values of the parameter
% $\mu > 0$. 

% It is also straightforward to incorporate flaky quantizers into this analysis.
% Then the result is that there exist a range of values for $\mu > 0$ 
% and $\nu_1 < \nu_2$. 

% It is even possible to allow for imprecise values of $\alpha$ and derive
% the admissible range for the full parameter set 
% $(\mu, \alpha, \nu_1, \nu_2)$. We omit this for now.

\subsection{Effect of additive noise and arithmetic errors on reconstruction
error} \label{GRE-4}

\begin{figure}[t]
\centerline{\includegraphics[width=4in]{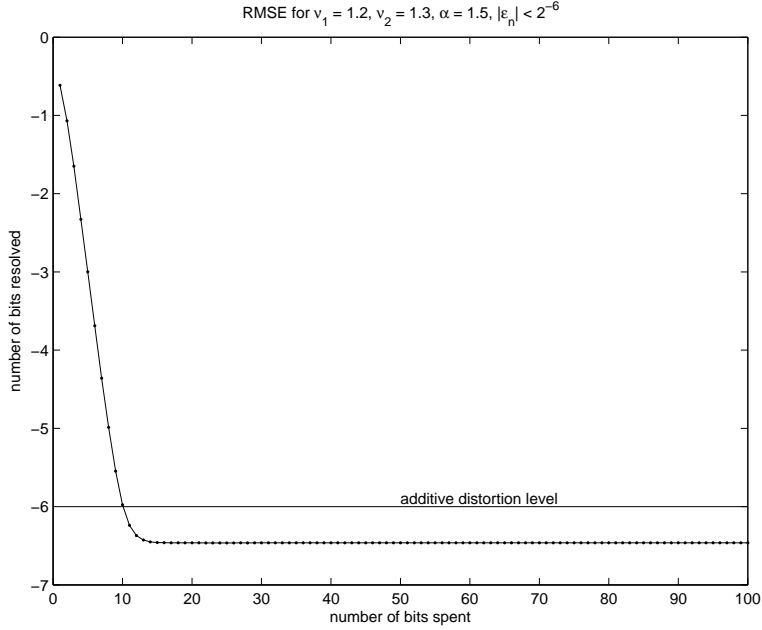}}
\caption{
\label{rec_error}
Demonstration of the fact that reconstruction error saturates at the
noise level. Here the parameters are $\nu_1 = 1.2$, $\nu_2 = 1.3$, 
$\alpha = 1.5$, and $|\epsilon_n| < 2^{-6}$.}
\end{figure}

Corollary \ref{cor_main} shows that the GRE is robust with respect to
quantizer imperfections under the assumption that the recursion
relations given by (\ref{rec}) are strictly satisfied. That is, we
have quite a bit of freedom in choosing the quantizer $Q$, assuming
the arithmetic, i.e., addition, can be done error-free. We now
investigate the effect of arithmetic errors on the reconstruction
error. To this end, we model such imperfections as additive
noise and, as before, replace (\ref{rec}) with (\ref{rec-n}),
where $\epsilon_n$ denotes the additive noise. 
% \begin{eqnarray}
% u_{n+2}&=&u_{n+1}+u_n-b_n+\epsilon_n \nonumber \\
% b_n &=&Q(u_n,u_{n+1}) \label{rec-n}
% \end{eqnarray}
While Theorem \ref{stab_thm} shows that the encoder is stable under
small additive errors, the reconstruction error is not guaranteed to
become arbitrarily small with increasing number of bits. This is
observed in Figure \ref{rec_error} where the system is stable for the
given imperfection parameters and noise level, however the
reconstruction error is never better than the noise level.

Note that for stable systems, we unavoidably have
\begin{equation} \label{err-n}
\sum_{n=0}^{N-1} b_n \phi^{-n}  = 
x + \sum_{n=0}^{N-1} \epsilon_n \phi^{-n} + O(\phi^{-N}),
\end{equation}
where the ``noise term'' in (\ref{err-n}) does not vanish as 
$N$ tends to infinity. To see this, define 
$\varepsilon_N:=\sum_{n=0}^{N-1} \epsilon_n \phi^{-n}$. 
If we assume $\epsilon_n$ to be i.i.d. with mean 0 and 
variance $\sigma^2$, then we have
$$\text{Var}(\varepsilon_N)=\frac{1-\phi^{-2N}}{1-\phi^{-2}} \sigma^2
\to \phi \,\sigma^2 \ \ \ \text{as}\ \ N\to \infty.$$ 
Hence we can conclude from this the $x$-independent result 
$$ \mathbf{E} \, \Big |x - \sum_{n=0}^\infty b_n \phi^{-n} \Big|^2 = \phi \, 
\sigma^2.$$

In Figure \ref{rec_error}, we incorporated uniform noise in the range
$[-2^{-6},2^{-6}]$. This would yield $\phi \sigma^2 = 2^{-12}\phi /3
\approx 2^{-13}$, hence the saturation of the root-mean-square-error
(RMSE) at $2^{-6.5}$. Note, however, that the figure was created with
an average over 10,000 randomly chosen $x$ values.  Although
independent experiments were not run for the same values of $x$, the
$x$-independence of the above formula enables us to predict the
outcome almost exactly.

In general, if $\rho$ is the probability density function for 
each $\epsilon_n$, then $\varepsilon_N$ will converge to a random variable
the probability density function of which has Fourier transform given by 
the convergent infinite product
$$ \prod_{n=0}^\infty \widehat \rho(\phi^{-n} \xi). $$

\subsection{Bias removal for the decoder} \label{GRE-5}

Due to the nature of any `cautious' beta-encoder, the standard $N$-bit
decoder for the GRE yields approximations that are biased, i.e., the
error $e_N(x)$ has a non-zero mean.  This is readily seen by the error
formula
$$ e_N(x) = \phi^{-N}(u_N + \phi u_{N+1}) $$
which implies that $e_N(x) > 0$ for all $x$ and $N$. Note that all
points $(u,v)$ in the invariant rectangle $R_Q$ satisfy $u + \phi v >
0$.

This suggests adding a constant ($x$-independent) term $\xi_N$ to the
standard $N$-bit decoding expression to minimize $\|e_N \|$. Various
choices are possible for the norm. For the $\infty$-norm, $\xi_N$
should be chosen to be average value of the minimum and the maximum
values of $e_N(x)$. For the $1$-norm, $\xi_N$ should be the median
value and for the $2$-norm, $\xi_N$ should be the mean value of
$e_N(x)$. Since we are interested in the $2$-norm, we will choose
$\xi_N$ via
$$\xi_N = \phi^{-N} \frac{1}{|I|} \int_I \left( u_N(x) + \phi\,u_{N+1}(x)
\right) \; dx,$$ where $I$ is the range of $x$ values and we have
assumed uniform distribution of $x$ values.

This integral is in general difficult to compute explicitly due to the
lack of a simple formula for $u_N(x)$. One heuristic that is motivated
by the mixing properties of the map $T_Q$ is to replace the average
value $|I|^{-1} \int_I u_N(x)\,dx$ by $\int_\Gamma u\,dudv$. If the
set of initial conditions $\{(x,0) : x\in I\}$ did not have zero
two-dimensional Lebesgue measure, this heuristic could be turned into
a rigorous result as well.

However, there is a special case in which the bias can be computed
explicitly.  This is the case $\alpha = 1$ and $\nu_1 = \nu_2 = 1$.
Then the invariant set is $[0,1)^2$ and
$$ \begin{bmatrix} u_N(x) \\ u_{N+1}(x) \end{bmatrix}
= \left \langle {\begin{bmatrix} 0 & 1 \\ 1 & 1 \end{bmatrix}}^N 
\begin{bmatrix} x \\ 0 \end{bmatrix}
\right \rangle
$$
where $\langle \cdot \rangle$ denotes the coordinate-wise fractional
part operator on any real vector.  Since $\int_0^1 \langle kx \rangle
\, dx = 1/2$ for all non-zero integers $k$, it follows that $\int_0^1
u_N(x) \, dx = 1/2$ for all $N$. Hence setting $I = [0,1]$, we find
$$\xi_N = \phi^{-N} \frac{1+\phi}{2} = \frac{1}{2} \phi^{-N+2}.$$

It is also possible to compute the integral $\int_\Gamma u\,dudv$
explicitly when $\alpha = \phi$ and $\nu_1 = \nu_2 = \nu$ for some
$\nu = [1,\phi]$.  In this case it can be shown that the invariant set
$\Gamma$ is the union of at most 3 rectangles whose axes are parallel
to $\Phi_1$ and $\Phi_2$. We omit the details.

\subsection{Circuit implementation: Requantization} \label{GRE-6}

As we mentioned briefly in the introduction, A/D converters (other
than PCM) typically incorporate a requantization stage after which the
more conventional binary (base-2) representations are generated. This
operation is close to a decoding operation, except it can be done
entirely in digital logic (i.e., perfect arithmetic) using the
bitstreams generated by the specific algorithm of the converter.  In
principle sophisticated digital circuits could also be employed.

\begin{figure}[t]
\centerline{\includegraphics[width=5in]{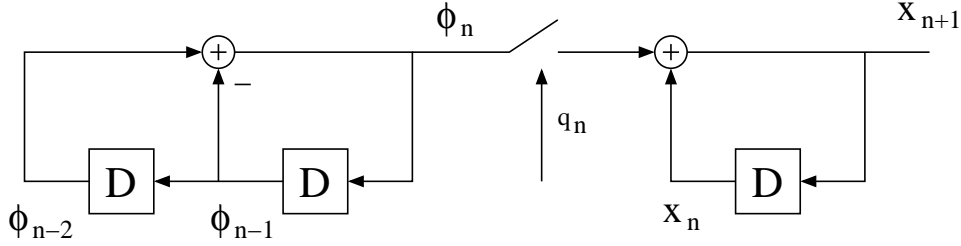}}
%\centerline{\includegraphics[width=3in]{beta2bin}}
\caption{
\label{beta2bin}
Efficient digital 
implementation of the requantization step for the Golden
Ratio Encoder.}
\end{figure}

In the case of the Golden Ratio Encoder, it turns out that a fairly
simple requantization algorithm exists that incorporates a digital
arithmetic unit and a minimum amount of memory that can be
hardwired. The algorithm is based on recursively computing the base-2
representations of powers of the golden ratio. In Figure
\ref{beta2bin}, $\phi_n$ denotes the $B$-bit base-2 representation of
$\phi^{-n}$. Mimicking the relation $\phi^{-n} = \phi^{-n+2} -
\phi^{-n+1}$, the digital circuit recursively sets
$$ \phi_n = \phi_{n-2} - \phi_{n-1}, \;\;\;\;\;\;n=2,3,...,N-1$$
which then gets multiplied by $q_n$ and added to $x_{n}$, where
$$x_n = \sum_{k=0}^{n-1} q_k \phi^{-k}, \;\;\;\;\;n=1,2,...,N.$$
The circuit needs to be set up so that $\phi_1$ is the expansion of
$\phi^{-1}$ in base 2, accurate up to at least $B$ bits, in addition to
the initial conditions $\phi_0 = 1$ and $x_0 = 0$.  To minimize
round-off errors, $B$ could be taken to be a large number (much larger
than $\log N/\log \phi$, which determines the output resolution).

\section{Higher order schemes: Tribonacci and Polynacci encoders} \label{poly}

What made the Golden Ratio Encoder (or the `Fibonacci' Encoder)
interesting was the fact that a beta-expansion for $1< \beta <2$ was
attained via a difference equation with $\pm 1$ coefficients, thereby
removing the necessity to have a perfect constant multiplier.  (Recall that 
multipliers were still
employed for the quantization operation, but they no longer needed
to be precise.)

This principle can be further exploited by considering more general
difference equations of this same type. An immediate class of such
equations are suggested by the recursion
$$ P_{n+L} = P_{n+L-1} + \dots + P_{n} $$ where $L > 1$ is some
integer. For $L=2$, one gets the Fibonacci sequence if the initial
condition is given by $P_0 = 0$, $P_1 = 1$. For $L=3$, one gets
the Tribonacci sequence when $P_0 = P_1 = 0$, $P_2 = 1$. The general
case yields the Polynacci sequence.

For bit encoding of real numbers, one then sets up the iteration
\begin{eqnarray}
u_{n+L}&=&u_{n+L-1}+\dots + u_n -b_n \nonumber \\
b_n &=&Q(u_n,\dots,u_{n+L-1})
\end{eqnarray}
with the initial conditions $u_0 = x$, $u_1 = \dots = u_{L-1} = 0$.
In $L$-dimensions, the iteration can be rewritten as

\begin{equation}
\begin{bmatrix*} u_{n+1} \\ u_{n+2} \\ \vdots \\ u_{n+L-1} \\ u_{n+L} \end{bmatrix*} 
= 
\begin{bmatrix} 
0 & 1 & 0 & \cdots & 0 \\
0 & 0 & 1 & \cdots & 0 \\
\vdots & \vdots & \ddots & \ddots & \vdots\\
0 & 0 & \cdots & 0 & 1 \\
1 & 1 & \cdots & 1 & 1
\end{bmatrix}
\begin{bmatrix*} u_{n} \\ u_{n+1} \\ \vdots \\ u_{n+L-2} \\ u_{n+L-1} \end{bmatrix*} 
-b_n \begin{bmatrix*} 0 \\ 0\\ \vdots \\ 0 \\ 1 \end{bmatrix*}
\end{equation}
It can be shown that the characteristic equation
$$ s^L - (s^{L-1} + \dots + 1) = 0 $$
has its largest root $\beta_L$ in the interval $(1,2)$ and all
remaining roots inside the unit circle (hence $\beta_L$ is a Pisot
number). Moreover as $L \to \infty$, one has $\beta_L \to 2$
monotonically.

One is then left with the construction of quantization rules $Q = Q_L$
that yield bounded sequences $(u_n)$. While this is a slightly more
difficult task to achieve, it is nevertheless possible to find such
quantization rules.  The details will be given in a separate
manuscript.

The final outcome of this generalization is the accuracy estimate
$$ \left |x - \sum_{n=0}^{N-1} b_n \beta_L^{-n} \right | = O(\beta_L^{-N})$$
whose rate becomes asymptotically optimal as $L \to \infty$.

\section{Acknowledgments}
We would like to thank Felix Krahmer, Rachel Ward, and Matt Yedlin for
various conversations and comments that have helped initiate and
improve this paper.

Ingrid Daubechies gratefully acknowledges partial support by the NSF
grant DMS-0504924.  Sinan G\"unt\"urk has been supported in part by
the NSF Grant CCF-0515187, an Alfred P. Sloan Research Fellowship, and
an NYU Goddard Fellowship. Yang Wang has been supported in part by the
NSF Grant DMS-0410062. {\"O}zg{\"u}r Y{\i}lmaz was partly supported by
a Discovery Grant from the Natural Sciences and Engineering Research
Council of Canada. This work was initiated during a BIRS Workshop and
finalized during an AIM Workshop. The authors greatfully acknowledge
the Banff International Research Station and the American Institute of
Mathematics. 
 
\frenchspacing

\bibliographystyle{IEEEtran}
\bibliography{GRE}

\end{document}